\documentclass[fleqn,usenatbib]{mnras}
\usepackage{newtxtext,newtxmath}
\usepackage[T1]{fontenc}
\usepackage[pdftex]{graphicx}

\title[Accretion by Proto-Jupiter in a Massive Planetesimal Disk]{Heavy-element Accretion by Proto-Jupiter in a Massive Planetesimal Disk, Revisited}

\author[Shibata et al.]{
S. Shibata,$^{1}$\thanks{E-mail: s.shibata423@gmail.com}
R. Helled,$^{1}$
and H. Kobayashi$^{2}$
\\
$^{1}$Institute for Computational Science (ICS), University of Zurich, Zurich, Switzerland\\
$^{2}$Department of Physics, Nagoya University, Japan\\
}

\date{Accepted 2022 November 29. Received 2022 November 25; in original form 2022 June 27}

\pubyear{2022}

\usepackage{amsmath}
\usepackage{comment}
\usepackage{color}

\newcommand{\g}{{\rm g}}

\newcommand{\cm}{{\rm cm}}
\newcommand{\m}{{\rm m}}

\newcommand{\AU}{{\rm ~au}}

\newcommand{\K}{{\rm K}}
\newcommand{\yr}{{\rm yr}}

\newcommand{\Msun}{M_{\odot}}

\newcommand{\Mear}{M_{\oplus}}

\begin{document}
\label{firstpage}
\pagerange{\pageref{firstpage}--\pageref{lastpage}}
\maketitle

\begin{abstract}
Planetesimal accretion is a key source for heavy-element enrichment in giant planets.
It has been suggested that Jupiter's enriched envelope is a result of planetesimal accretion during its growth assuming it formed in a massive planetesimal disk. 
In this study, we simulate Jupiter's formation in this scenario.  
We assume in-situ formation and perform N-body simulations to infer the solid accretion rate. 
We find that tens-Earth masses of planetesimals can be captured by proto-Jupiter during the rapid gas accretion phase. 
However, if several embryos are formed near  Jupiter's core, which is an expected outcome in the case of a massive planetesimal disk,  scattering from the embryos increases the eccentricity and inclination of planetesimals and therefore significantly reduces the  accretion efficiency. 
We also compare our results with published semi-analytical models and show that these models cannot reproduce the N-body simulations especially when the planetesimal disk has a large eccentricity and inclination.
We show that when the dynamical evolution of planetesimals is carefully modelled, the total mass of captured planetesimals $M_\mathrm{cap,tot}$ is $2 M_\oplus \lesssim M_\mathrm{cap,tot}\lesssim 18 M_\oplus$.
The metallicity of Jupiter's envelope can be explained by the planetesimal accretion 
in our massive disk model despite the low accretion efficiency coming from the high eccentricity and inclination of planetesimals. 
Our study demonstrates the importance of detailed modelling of planetesimal accretion during the planetary growth and its implications to the heavy-element mass in gaseous planets. 
\end{abstract}


\section{Introduction} \label{sec:intro}
{
Understanding the origin of the heavy-element mass in Jupiter is important for giant planet formation theory and for constraining the conditions of the protoplanetary disk from which the solar system formed. 
The origin of the heavy-element enrichment in Jupiter's atmosphere remains unknown. 
The bulk composition of Jupiter is not well determined, and currently the estimated heavy-element mass is between $10$ and $45 \Mear$ \citep[e.g.][]{Wahl+2017,Debras+2019,Nettelmann+2021,Miguel+2022,Helled+2022}. 

In the classic core accretion model, the heavy-element core is formed as a result of planetesimal accretion. % is a fundamental process for the planetary core formation.
Once the growing core reaches a critical mass, the planetary core enters the runaway gas accretion phase and a giant planet is formed \citep[e.g.][]{Mizuno+1980}.
Planetesimal accretion is expected to continue even during rapid gas accretion and contributes to the overall enrichment of the planetary envelope. 
Pebble accretion is an alternative mechanism to form the planetary core \citep[e.g.][]{Lambrechts+2012}, however, pebble accretion is then halted once the pebble isolation mass has been reached \citep{Lambrechts+2014,Bitsch+2018}.
In this work we focus on planetesimal accretion during runaway gas accretion and determine the expected enrichment of Jupiter's envelope.

The planetesimal accretion rate during the runaway gas accretion phase can be investigated by the N-body simulations around the growing protoplanet  \citep[e.g.,][]{Zhou+2007,Shiraishi+2008,Shibata+2019,Podolak+2020,Eriksson+2022}.
A massive protoplanet opens a gap in the gas disk and migrates in the radial direction in a type II configuration \citep[e.g.][]{Kanagawa+2018}.
Planetary migration is found to enhance the planetesimal accretion rate since many planetesimals are supplied into the region of the protoplanet's orbit \citep[e.g.,][]{Alibert+2005,Shibata+2020,Shibata+2022a,Turrini+2021,Shibata+2022}. 
However, in order to reach a metallicity of $\sim$3 time solar in the planetary envelope, as expected for Jupiter, the planetesimal disk must have been several times more massive than the minimum mass solar nebulae (MMSN) \citep{Venturini+2020}.

Recently, \citet{Kobayashi+2021} suggested a mechanism to form a massive planetesimal disk in the inner solar system ($\lesssim10\AU$). 
They simulated the collisional evolution from dust to planets in the entire disk and showed that when using realistic porosity of dust aggregates, planetesimals are formed in $\sim 10$\AU. Planetesimals can also grow from pebbles that form in the outer disk and drift to the inner disk. As a result, 
the solid surface density can reach $20\g/\cm^2$ around $\sim6\AU$ and a planetary core of $10M_{\oplus}$ forms within $2\times10^5 \yr$.
If the planetary core enters the runaway gas accretion phase there, the planetesimal accretion rate could be so high that tens Earth-masses of heavy elements are captured by the end of Jupiter's formation. 

In the scenario of a massive planetesimals disk, however, the orbits of the planetesimals would be excited because of the mutual excitation \citep[e.g.][]{Ohtsuki+2002}.
If there are other embryos around Jupiter's core, the orbits of planetesimals would be excited even further. 
In previous studies, the assumed solid surface density was close to that of the MMSN. In this case, the mutual gravitational scattering of planetesimals are negligibly small relative to that from proto-Jupiter, and the existence of other embryos is ignored.
Therefore, the initial eccentricity and inclination of planetesimals are set as small as or smaller than $10^{-3}$ \citep{Zhou+2007, Shiraishi+2008, Shibata+2019, Podolak+2020}.
However, if the planetesimal disk is massive, planetesimals would be excited and the capture efficiency would decrease significantly \citep[e.g.][]{Inaba+2001,Chambers2006,Fortier+2013}.
%The captured mass of planetesimals would be also reduced from the results obtained in previous studies.

In this study, we revisit the scenario of planetesimal accretion onto the growing proto-Jupiter in a massive planetesimal disk. % assuming the massive planetesimal disk
Our model updates from previous studies in three main points;
i) the distribution of planetesimals, % \citet{Kobayashi+2021}
ii) Jupiter's formation model including the effects of gap formation and planetary migration,
iii) initial orbital elements of planetesimal disk.
We perform N-body simulations with the step-by-step updates.
Also, we compare our results with the semi-analytical models found in literatures.
In Sec.~\ref{sec:model}, we describe the method and model in our study.
In Sec.~\ref{sec:planetesimal_accretion}, we show the numerical results.
As already found in previous studies \citep{Zhou+2007,Shibata+2020,Shibata+2022a}, mean motion resonances of proto-Jupiter play important roles in the planetesimal accretion process.
We also analyse the numerical results focusing on the mean motion resonances in Sec.~\ref{sec:MMRs}.
Section~\ref{sec:comparison} is dedicated to the comparison of our results with the semi-analytical models.
We compare our results with the observations in Sec.~\ref{sec:discussion}.
We also discuss the phase 2 there.
A summary of the work is given in  Sec.~\ref{sec:Conclusion}.
}

\section{Methods} \label{sec:model}
{
%\subsection{The equation of motion}
In this study we use the orbital integration code presented in \citet{Shibata+2019}. 
We perform orbital integrations of a central star with mass $M_{*}$, a growing proto-Jupiter with mass $M_{\rm p}$, and a massive planetesimal disk.
Our simulations begin when proto-Jupiter enters the runaway gas accretion phase.
The runaway gas accretion is defined as the phase where the mass of the planetary core  $M_\mathrm{core}$ is larger than the critical core mass $M_{\rm crit}$ and the mass of the planetary envelope exceeds $M_\mathrm{core}$ (also known as cross-over mass $M_{\rm cross}$).
In this definition, the initial mass of proto-Jupiter $M_\mathrm{p,0}$ is given as $2M_{\rm cross}$.
The planet enters the detached phase once the gas accretion rate supplied by the protoplanetary disk is smaller than the rate required to prevent envelope contraction. This phase continues until the protoplanetary disk dissipates.

During gas accretion, proto-Jupiter migrates inward due to the tidal interaction with the surrounding gaseous disk. 
The gas accretion and the planetary migration models of proto-Jupiter are described in sec.~\ref{sec:method_EvPath}.
The surrounding planetesimals feel gas drag from the gaseous disk, and we adopt the gas drag model of \citet{Adachi+1976}.
We assume the vertically isothermal disk and the disk gas rotates with the sub-Kepler velocity.
The velocity and density of the ambient disk gas are calculated from the protoplanetary disk model (see sec.~\ref{sec:method_disk}).
Below we summarise the model, further details on the simulations can be found in  \citet[][]{Shibata+2019, Shibata+2022}, and in appendix~\ref{app:method}. 

\subsection{Evolution pathways of growing gas giant planets}\label{sec:method_EvPath}
We adopt the model of planetary migration presented by \citet{Kanagawa+2018}.
Performing hydro-dynamical simulations with various disk parameters, \citet{Kanagawa+2018} found that the planetary migration rate can be scaled with the surface density of disk gas at the gap bottom opened by the protoplanet $\Sigma_{\rm gap}$ and derived the migration rate as:
\begin{align}
    \frac{d \ln r_{\rm p}}{d t} &=  -2 C_\mathrm{M} \frac{M_{\rm p}}{M_{*}}
                                \frac{{r_{\rm p}}^2\Sigma_{\rm gap}}{M_{*}}
                                \left(\frac{h_{\rm s}}{r_{\rm p}} \right)^{-2} \Omega_{\rm p}, \label{eq:type2_migration_K18}
\end{align} 
where $C_\mathrm{M}$ is a factor that depends on Lindblad and corotation torques, 
$r_{\rm p}$ is a radial distance of the protoplanet from the central star,
$\Sigma_\mathrm{gap}$ is a surface density of disk gas at the gap bottom,
$h_\mathrm{s}$ is a scale height of the disk gas, and
$\Omega_{\rm p}$ is the Kepler angular velocity of the protoplanet.
In this model, the migration mode smoothly shifts from type I regime to type II regime as the protoplanet opens a gap in the disk.

For gas accretion, we adopt the model obtained in \citet{Tanigawa+2002}.
\citet{Tanigawa+2002} found that accreting gas onto the protoplanet passes through the narrow band region and that the gas accretion rate is regulated by the width of the accretion band and the speed of gas flow at the band region.
They derived the following empirical formula for the gas accretion rate: 
\begin{align}
    \frac{d M_{\rm p}}{d t} = D \Sigma_{\rm gap} \label{eq:GasAccretionRate_TT}
\end{align}
with
\begin{align}
    D = 0.29 \left( \frac{M_{\rm p}}{M_{*}} \right)^{4/3} \left( \frac{h_{\rm p}}{r_{\rm p}} \right)^{-2} {r_{\rm p}}^2 \Omega_{\rm p}.
\end{align}

As pointed by \citet{Tanaka+2020}, when the planetary mass is similar to the critical core mass, a slow Kelvin-Helmholtz contraction of the planetary envelope regulates the gas accretion and eq.~(\ref{eq:GasAccretionRate_TT}) overestimates the gas accretion rate. 
To account for the slow contraction, we constrain the upper boundary of the gas accretion timescale using Kelvin-Helmholtz timescale given by:  \citep[e.g.][]{Ikoma+2000, Ida+2018}
\begin{align}\label{eq:Kelvein-Helmholz_timescale}
    \tau_{\rm KH} = 1 \times10^3 \yr \left( \frac{M_{\rm p}}{100 M_{\oplus}} \right)^{-3.0} \left( \frac{\kappa}{1 \cm^2 \g^{-1}} \right),
\end{align}
where $\kappa$ is the opacity of the planetary envelope. 
In this study, we consider a grain depleted case and set  $\kappa=0.05 \cm^2 \g^{-1}$.
Even if we use the higher opacity such as $\kappa=1 \cm^2 \g^{-1}$, the gas accretion regime shifts from the attached phase (eq.~(\ref{eq:Kelvein-Helmholz_timescale})) to the detached phase (eq.~(\ref{eq:GasAccretionRate_TT})) before reaching $M_{\rm p}=30M_{\oplus}$ (see fig.~\ref{fig:mp_timescales}), and then the gas accretion rate becomes independent of the opacity.
Thus, the effect of the opacity is limited to the early gas accretion phase. 
The evolution pathway of the protoplanet is mainly controlled by eq.~(\ref{eq:type2_migration_K18}) and eq.~(\ref{eq:GasAccretionRate_TT}).

From eq.~(\ref{eq:type2_migration_K18}) and eq.~(\ref{eq:GasAccretionRate_TT}), the fraction of migration timescale $\tau_{\rm tide,{\it a}}$ and gas accretion timescale $\tau_{\rm acc}$ is given by: 
\begin{align}\label{eq:fraction_of_timescales}
    \frac{\tau_{\rm tide,{\it a}}}{\tau_{\rm acc}} \sim  \frac{15}{\left|C_{\rm M}\right|} \left( \frac{M_{\rm p}}{M_{\rm J}} \right)^{-2/3}.
\end{align}
$C_{\rm M}$ is given as a summation of the normalised Lindblad and corotation torques (see eq.~(29) in \citet{Kanagawa+2018}) and depends on the local temperature and density gradient, but calculating exact value of $C_{\rm M}$ is beyond the scope of this study.
\citet{Tanaka+2020} found that if $C_{\rm M}$ is independent of the disk structure and can be set as a constant, the evolution pathways on $a_{\rm p}$-$M_{\rm p}$ plane is independent of the disk's profile $\Sigma_{\rm gas}$. 
We follow the model of \citet{Tanaka+2020} and set $C_{\rm M}=2$.
Equation (\ref{eq:fraction_of_timescales}) means that during Jupiter's formation ($M_{\rm p}\leq~M_{\rm J}$), $\tau_{\rm tide,{\it a}}$ is much larger than $\tau_{\rm acc}$.
Therefore, proto-Jupiter barely migrates in radial direction (less than $2\AU$ in our model) before reaching Jupiter's mass.
Using this nearly in-situ formation scenario for Jupiter, we investigate the efficiency of planetesimal accretion onto its envelope during gas accretion.

Note that our formation model begins at the onset of the runaway gas accretion and does not include the formation of the core. 
Jupiter's core could have migrated several au in the radial direction via type I migration, and grown via planetesimal accretion. 
The planetesimals captured when $M_{\rm p} < M_\mathrm{p,0}$ are expected to  join Jupiter's core and are not considered in this study.
The total heavy-element mass in Jupiter $M_\mathrm{Z,Jup}$ includes the initial core mass $M_\mathrm{core,0}$ and the captured heavy-element mass after the onset of runaway gas accretion $M_{\rm cap,tot}$.
In this study, we derive $M_{\rm cap,tot}$ below and discuss the uncertainty in $M_\mathrm{core,0}$ in sec.~\ref{sec:discussion_Mp0}.

\subsection{Disk model}\label{sec:method_disk} 
Our disk model is based on the model developed in \citet{Shibata+2022}.
The surface density profile of the gaseous disk is given by the self-similar solution \citep{Lynden-Bell+1974}.
We also include the effects of the gap opening, the feedback from the gas accretion and the disk depletion.
Thus, the surface density of disk gas evolves with time and depends on the disk viscosity $\alpha_{\rm acc}$ and the disk depletion timescale $\tau_{\rm dep}$. 

%%%%%%%%%%%%%%%%%%%%%%%%%%%%%%%%%%%%%%%%%%%%%%%%%%%%%%%%%%%%%
\begin{figure}
  \begin{center}
    \includegraphics[width=90mm]{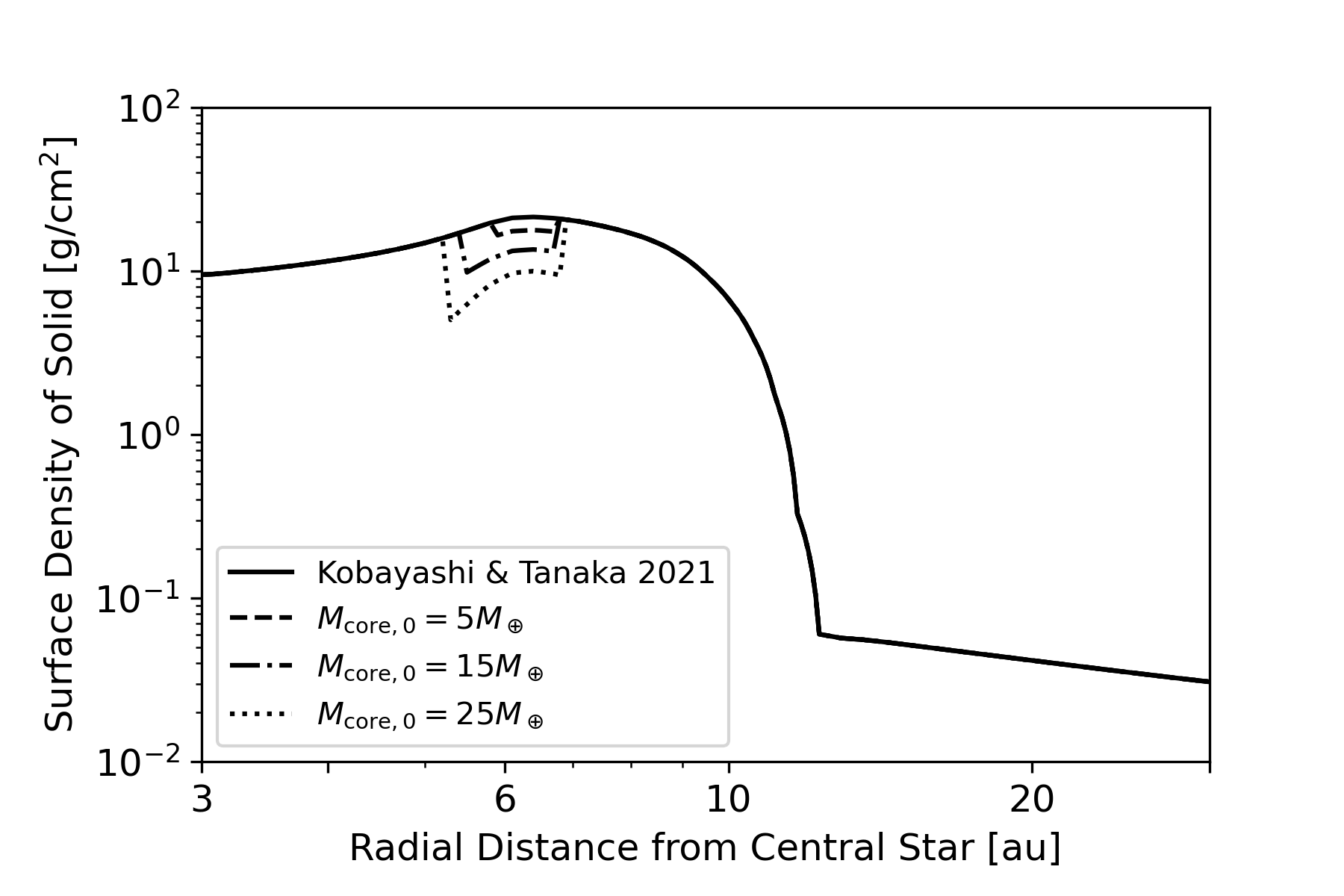}
    \caption{
    Surface density profile of planetesimals we use in this study.
    The solid line shows the surface density of solids obtained in \citet{Kobayashi+2021}.
    The dashed, dash-dotted, and dotted lines are $\Sigma_\mathrm{solid}$ with $M_\mathrm{core,0}=5 M_\oplus$, $15 M_\oplus$, and $25 M_\oplus$, respectively.
    }
    \label{fig:disk_solid}
  \end{center}
\end{figure}
%%%%%%%%%%%%%%%%%%%%%%%%%%%%%%%%%%%%%%%%%%%%%%%%%%%%%%%%%%%%%

The planetesimal disk is usually set to have a power-law radial distribution, similar to the solid distribution in protoplanetary disks. 
However, \citet{Kobayashi+2021} simulated the collisional evolution of dust grains in  the disk, and found that this leads to a configuration of a dense-compact planetesimal disk due to pebble drift from the outer disk. 
Figure \ref{fig:disk_solid} shows the solid surface density profile $\Sigma_\mathrm{solid}$ at $\sim1.9 \times 10^5$ years inferred by \citet{Kobayashi+2021} where a core of $10 M_\oplus$ is formed at $\sim 6~\AU$. 
\citet{Kobayashi+2021} shows that planetesimals are main mass reservoir after a core is formed.

At the beginning of the simulations, we assume that Jupiter's core of $M_\mathrm{core,0}$ is already formed.
In order to consider the decrease in the available heavy-element mass for accretion around the protoplanet due to core formation, we reduce the surface density of planetesimals by $M_\mathrm{core,0}$.
The exact formula of $\Sigma_\mathrm{solid}$ is given in appendix~\ref{app:method_planetesimal_disk}.

We follow the orbital motion of super-particles, each of which contains several equal-size planetesimals.
The super-particles are treated as test particles, thus the mutual gravity is neglected during the calculation.
The super-particles are distributed radially along the planetary feeding zone where proto-Jupiter could accrete them as it grows. 
We set the number of super particles $N_{\rm sp}=9600$ where the spatial density is kept larger than $2,000$ super-particles per $1\AU$.  

Assuming that planetesimals have been scattered by their mutual gravitational interactions, we adopt the Rayleigh distributions for the initial eccentricities $e$ and inclinations $i$ of the planetesimals. 
The initial root-mean-square values of eccentricities $\langle {e_0}^2\rangle^{1/2}$ and inclinations $\langle \sin^2 i_0 \rangle^{1/2}$ are set as input parameters.
The other orbital angles, such as the longitude of ascending node $\Omega$, the argument of perihelion $\omega$, and the mean longitude at epoch $\epsilon$ are distributed uniformly.

\subsection{Parameter settings}
%%%%%%%%%%%%%%%%%%%%%%%%%%%%%%%%%%%%%%%%%%%%%%%%%%%%%%%%%%%%%
\begin{table*}
%{\scriptsize
	\centering
	\begin{tabular}{ llllll } % four columns, alignment for each
		\hline
		\hline
		$M_{*}$ 		& Mass of central star	                        & $1.0 \Msun$ &	\\
		$M_{\rm disk,0}$    & Initial mass of protoplanetary disk           & $0.037 \Msun$ &	\\
		$R_{\rm disk}$      & Typical size of protoplanetary disk           & $108 \AU$ &	\\
		$T_{\rm disk,0}$    & Disk mid-plane temperature at $1\AU$           & $200 \K$ &	\\
		$\tau_{\rm dep}$ 	& Disk depletion timescale                      & $1\times10^6 \yr$ & \\
		$\rho_{\rm p}$ 	    & Mean density of protoplanet                   & $0.125 \g/\cm^{-3}$ & \\
		$\rho_{\rm pl}$ 	& Mean density of planetesimals                 & $1.0 \g/\cm^{-3}$ & \\
		$N_{\rm sp}$ 	    & Number of super-particles                     & $9600$ & \\
%		\hline
%		$M_\mathrm{core}$ & Mass of Jupiter's core                & &	\\
%		$M_\mathrm{core,0}$ & Initial mass of Jupiter's core                & &	\\
%		$M_\mathrm{crit}$ & Critical core mass                & &	\\
%		$M_\mathrm{cap}$    & Cumulative mass of captured planetesimals     & &	\\
%		$M_\mathrm{cap,tot}$    & Total mass of captured planetesimals     & &	\\
%		$M_\mathrm{Z,Jup}$    & Total mass of heavy elements in Jupiter    & &	\\
%		$R_{\rm pl}$ 	    & Radius of planetesimals                       & $10^7, 10^6, 10^5\cm$ & \\
%		$M_{\rm p,0}$ 		& Initial mass of protoplanet	                & $3\times10^{-5} \Mear$ & \\
%		$\rho_{\rm p}$      & Mean density of protoplanet                   & $0.125 g/\cm^3$ & \\
		\hline
	\end{tabular}
	\caption{
	Parameters used in our simulations.
    }
    \label{tb:model_settings1}
%}
\end{table*}
%%%%%%%%%%%%%%%%%%%%%%%%%%%%%%%%%%%%%%%%%%%%%%%%%%%%%%%%%%%%%

%%%%%%%%%%%%%%%%%%%%%%%%%%%%%%%%%%%%%%%%%%%%%%%%%%%%%%%%%%%%%
\begin{table*}
%{\scriptsize
	\centering
	\begin{tabular}{ llllll } % four columns, alignment for each
		\hline
		\hline
		                    &  & Model-A & Model-B \\ %& Case 2 \\
		\hline
		$\tau_{\mathrm{tide},a}$ & Migration timescale                   & $\infty$ & Eq.~(\ref{eq:type2_migration_K18}) \\ %& Eq.~(\ref{eq:type2_migration_K18}) \\
		$M_{\rm p,0}$ 		& Initial planetary mass                    & $10 M_{\oplus}$ & $10, 30, 50 M_{\oplus}$ \\ %& $10 M_{\oplus}$  \\
		$a_{\rm p,0}$ 		& Initial semi-major axis of protoplanet    & $5.2 \AU$ & $6.3 \AU$ \\ %& $6.5 \AU$  \\
		$t_{\rm 0}$ 		& Formation time of planetary core		    & $2.4\times10^5 \yr$ & $2.4\times10^5 \yr$ \\ %& $1.9\times10^5 \yr$ \\
		$\alpha_{\rm acc}$ 	& Disk accretion viscosity		            & $6.3\times10^{-4}$ & $6.3\times10^{-4}$ \\ %& $1.0\times10^{-3}$ \\
		\hline
		$R_{\rm pl}$ 	    & Radius of planetesimals                   & $10^5$-$10^8\cm$ & $10^5$-$10^8\cm$ \\
		$\langle {e_0}^2\rangle^{1/2}$ 	& Mean square value of eccentricities       & $10^{-3}$ & $10^{-3}$-$0.4$ \\
		$\langle \sin^2 i_0 \rangle^{1/2}$ 	& Mean square value of inclinations         & $0.5 \times 10^{-3}$ & $0.5 \times 10^{-3}$-$0.2$ \\
		\hline
	\end{tabular}
	\caption{
	Parameters used in each formation models.
    }
    \label{tb:model_settings2}
%}
\end{table*}
%%%%%%%%%%%%%%%%%%%%%%%%%%%%%%%%%%%%%%%%%%%%%%%%%%%%%%%%%%%%%

%%%%%%%%%%%%%%%%%%%%%%%%%%%%%%%%%%%%%%%%%%%%%%%%%%%%%%%%%%%%%
\begin{figure}
  \begin{center}
    \includegraphics[width=80mm]{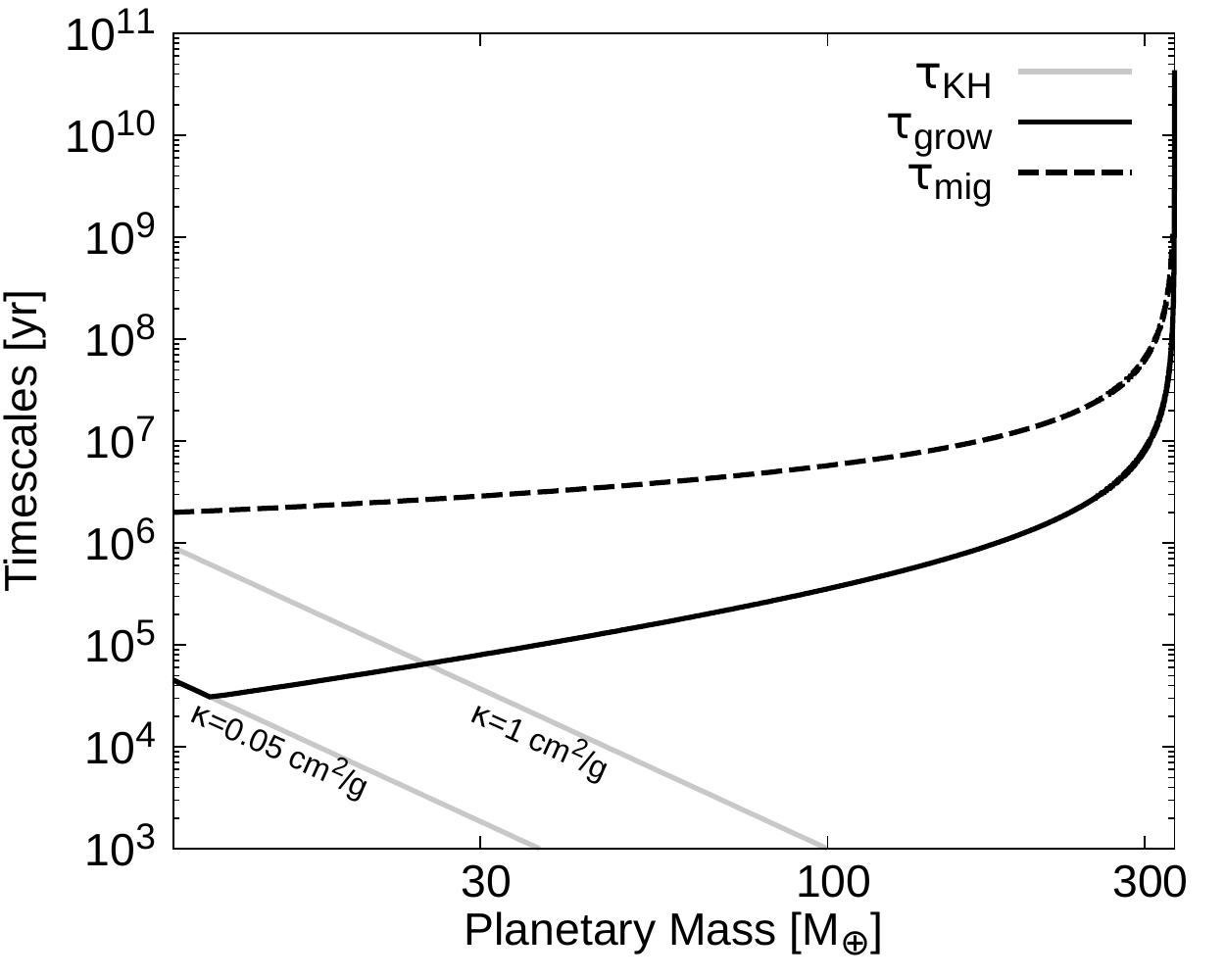}
    \caption{
    Gas accretion timescale $\tau_{\rm acc}$ (solid line) and migration timescale $\tau_{\mathrm{tide},a}$ (dashed line) as a function of planetary mass in Model-B. 
    The gray solid lines show the Kelvin-Helmholz timescale given by eq.~(\ref{eq:Kelvein-Helmholz_timescale}) with $\kappa=1 \cm^2/\g$ (upper line) and $\kappa=0.05 \cm^2/\g$ (lower line). 
    }
    \label{fig:mp_timescales}
  \end{center}
\end{figure}
%%%%%%%%%%%%%%%%%%%%%%%%%%%%%%%%%%%%%%%%%%%%%%%%%%%%%%%%%%%%%

We start the simulations with a protoplanet which enters runaway gas accretion with $a_{\rm p}=a_{\rm p,0}$ and $M_{\rm p}=M_{\rm p,0}$ at $t=t_0$. 
The mass and orbit of the protoplanet evolve according to the model introduced in Sec.~\ref{sec:method_EvPath}. 
The gas accretion rate and planetary migration rate become negligibly small when the disk gas depletes. 
The final semi-major axis $a_{\rm p,f}$ and the final mass $M_{\rm p,f}$ depend on $t_0$ and $\tau_{\rm dep}$ \citep{Tanaka+2020}.

In this study, we set $M_{\rm p,0}=3\times10^{-5} M_{*}\sim10 M_{\oplus}$ and $\tau_{\rm dep}=1\times10^6\yr$.
We also consider the cases with more massive core $M_{\rm p,0}=30$ and $ 50 M_{\oplus}$ in sec.~\ref{sec:discussion}.
Given $\alpha_{\rm acc}$, we can find the parameter set $(a_{\rm p,0}, t_0)$ that forms Jupiter with $M_{\rm p,f}=318 M_{\oplus}$ and $a_{\rm p,f}=5.2\AU$.
We consider two formation models: 
(i) Model-A, where we artificially neglect the effect of orbital migration where 
we set $d \ln r_{\rm p}/dt = 0$ instead of eq.~(\ref{eq:type2_migration_K18}). 
This is the same setting as previous studies \citep{Zhou+2007,Shiraishi+2008,Shibata+2019,Podolak+2020}.
To be consistent with the result of \citet{Kobayashi+2021} where Jupiter's core formed with $t_0\sim2\times10^5\yr$, we adapt $\alpha_{\rm acc}=6.3\times10^{-4}$ and obtain $t_0=2.4\times10^5\yr$.
(ii) Model-B, where we adopt the migration model given by eq.~(\ref{eq:type2_migration_K18}).
In this case, $\alpha_{\rm acc}$ and $t_0$ are $6.3\times10^{-4}$ and $2.4\times10^5\yr$, respectively.

In previous work we showed that the planetesimal accretion rate depends on the strength of aerodynamic gas drag \citep[e.g.][]{Shibata+2019}. 
To investigate this effect further, we perform parameter studies regarding the size of planetesimals $R_{\rm pl}$.
In addition, in previous studies the initial eccentricity $\langle {e_0}^2\rangle^{1/2}$ and the initial inclination $\langle \sin^2 i_0 \rangle^{1/2}$ are set as small values of $\lesssim10^{-3}$ \citep[e.g.,][]{Zhou+2007,Shiraishi+2008,Shibata+2019,Podolak+2020}.
However, the values of $\langle {e_0}^2\rangle^{1/2}$ and $\langle \sin^2 i_0 \rangle^{1/2}$ are determined by the viscous stirring between the planetesimals, and the strength of the viscous stirring increases with the surface density of planetesimals \citep[e.g.][]{Ohtsuki+2002}.
In a massive planetesimal disk, $\langle {e_0}^2\rangle^{1/2}$ and $\langle \sin^2 i_0 \rangle^{1/2}$ can be larger than $10^{-3}$.
In addition, if planetary embryos form in the planetesimal disk, the viscous stirring from the embryos 
significantly contributes to planetesimal dynamical excitation.
\citet{Kobayashi+2021} found that embryos as massive as Earth  or more form in the planetesimal disk near Jupiter's core. 
In order to include these effects we also perform the parameter study regarding $\langle {e_0}^2\rangle^{1/2}$ and $\langle \sin^2 i_0 \rangle^{1/2}$.

The parameters used in our simulations are listed in  tables.~\ref{tb:model_settings1} and \ref{tb:model_settings2}.
Figure~\ref{fig:mp_timescales} shows the evolution of timescales $\tau_{\rm acc}$ and $\tau_{\rm mig}$ as a function of planetary mass obtained in Model B. 
Other settings and parameters in this model, such as the capture radius of proto-Jupiter, are described in appendix~\ref{app:method}.
}

\section{Results} \label{sec:planetesimal_accretion}
{
\subsection{Planetesimal accretion onto a non-migrating proto-Jupiter}\label{sec:results_woMig}
%%%%%%%%%%%%%%%%%%%%%%%%%%%%%%%%%%%%%%%%%%%%%%%%%%%%%%%%%%%%%
\begin{figure*}
  \begin{center}
    \includegraphics[width=160mm]{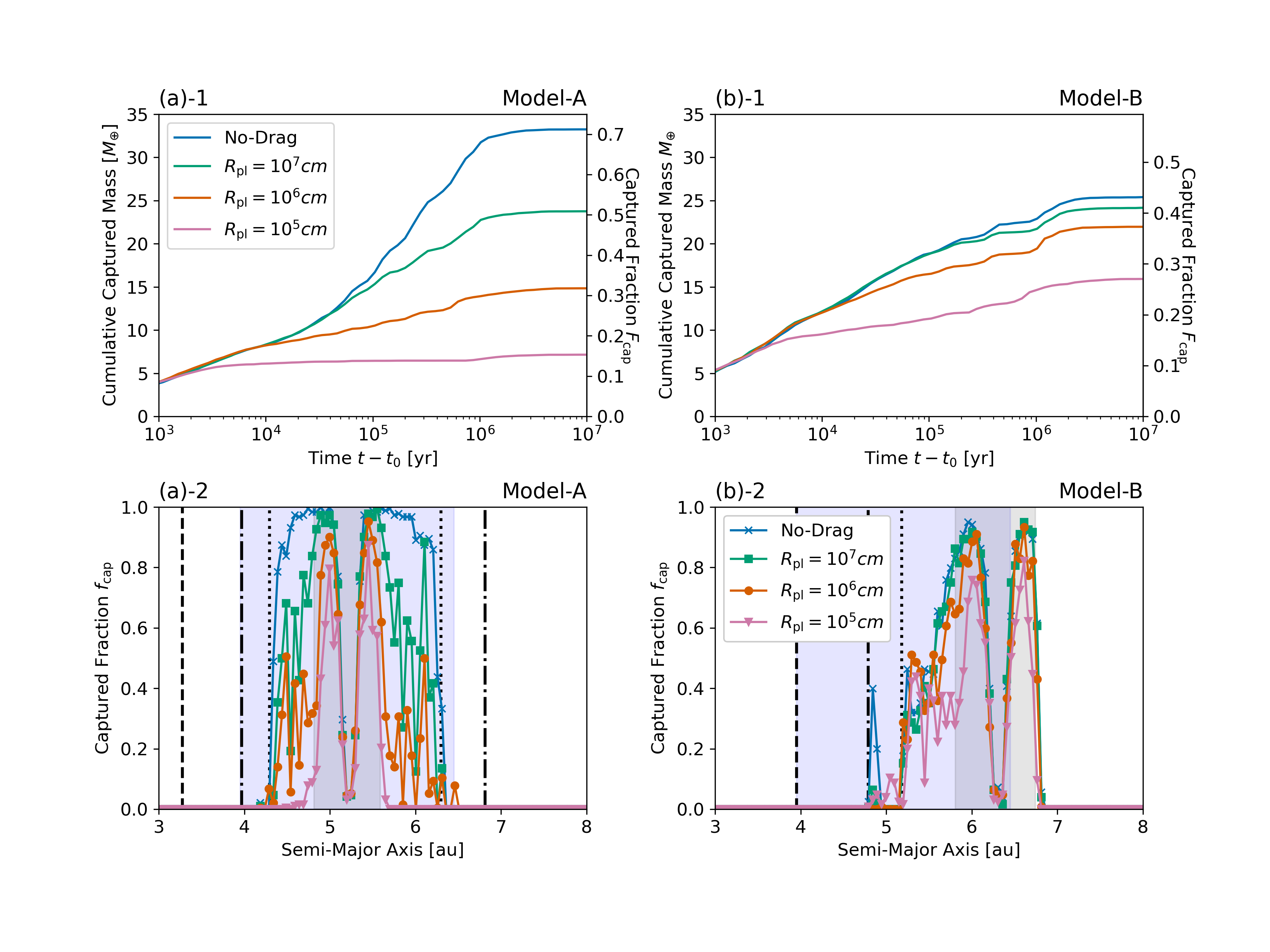}
    \caption{
    Results of our N-body simulations.
    Left column shows the results obtained in Model-A and right column shows the results obtained in Model-B.
    {\bf Upper panels}: 
    Change in the cumulative mass of captured planetesimals $M_\mathrm{cap}$ as a function of calculation time $t-t_0$.
    The green, orange and magenta lines show the cases of $R_{\rm pl}=10^{7}\cm$, $10^{6}\cm$ and $10^{5}\cm$, respectively.
    The blue line shows the case where the aerodynamic gas drag is artificially neglected.
    {\bf Lower panels}: 
    Fraction of captured planetesimals as a function of initial semi-major axes of planetesimals.
    The green, orange and magenta lines show the cases of $R_{\rm pl}=10^{7}\cm$, $10^{6}\cm$ and $10^{5}\cm$, respectively.
    The blue line shows the case where the aerodynamic gas drag is artificially neglected. 
    The blue shaded areas from $3.95\AU$ to $6.45\AU$ indicate the feeding zone at the end of the simulation.
    The gray shaded areas from $4.81\AU$ to $5.89\AU$ in Model-A and from $5.81\AU$ to $6.74\AU$ in Model-B,  represent the initial feeding zone. 
    The vertical black lines show the location of resonance centre at the beginning of the simulations for $2:1$ (dashed line), $3:2$ and $2:3$ (dash-dotted lines), and $4:3$ and $3:4$ (dotted line), respectively.
    }
    \label{fig:Results_psRpl}
  \end{center}
\end{figure*}
%%%%%%%%%%%%%%%%%%%%%%%%%%%%%%%%%%%%%%%%%%%%%%%%%%%%%%%%%%%%%

First, we show the results for Model-A where proto-Jupiter does not migrate.
Panel (a)-1 in fig.~\ref{fig:Results_psRpl} shows the change in the cumulative mass of planetesimals captured by proto-Jupiter $M_\mathrm{cap}$.
 
As found by previous studies, we find that proto-Jupiter can capture more planetesimals as the planetary mass increases.
In the right vertical axis of panel (a)-1, we plot the fraction of captured planetesimals $F_{\rm cap}$ which is defined by: 
\begin{align}\label{eq:definition_Fcap}
    F_{\rm cap} = \frac{M_{\rm cap}}{M_{\rm FZ,tot}},
\end{align}
where $M_{\rm FZ,tot}$ is the total mass of planetesimals %$M_{\rm cap}$ is the cumulative mass of captured planetesimals, and
expected to be swept by the protoplanet's feeding zone during runway gas accretion. 

The feeding zone is defined as a region where the Jacobi energy $E_{\rm Jacobi}$ is positive \citep[e.g.][]{Hayashi+1977}.
The Jacobi energy is given by \citep[e.g.][]{Hayashi+1977}: 
\begin{align}
    E_{\rm Jacobi} = \frac{\mathcal{G} M_{*}}{a_{\rm p}} \left\{ -\frac{a_{\rm p}}{2 a} - \sqrt{\frac{a}{a_{\rm p}} \left(1-e^2\right)} \cos {i} + \frac{3}{2} + \frac{9}{2} h^2 +O(h^3) \right\}, \label{eq:Jacobi_Energy}
\end{align}
where $a$ is the semi-major axis of a planetesimal, $\mathcal{G}$ is the gravitational constant and $h$ is the reduced Hill radius:  
\begin{align}\label{eq:reduced_Hill_radius}
    h = \left( \frac{M_{\rm p}}{3 M_{*}} \right)^{1/3}.
\end{align}
We define the normalized Jacobi energy as: % $\tilde{E}_{\rm Jacobi}  \equiv a_{\rm p}/\mathcal{G} M_{*} h^2 E_{\rm Jacobi}$.
%For the planetesimals around proto-Jupiter, eq.~(\ref{eq:Jacobi_Energy}) is approximated as
\begin{align}
    \tilde{E}_{\rm Jacobi}  &\equiv \frac{a_{\rm p}}{\mathcal{G} M_{*}} \frac{E_{\rm Jacobi}}{h^2}, \\
                            &\sim \frac{1}{2} \left( \tilde{e}^2 + \tilde{i}^2 \right) -\frac{3}{8} \tilde{b}^2 +\frac{9}{2}, \label{eq:Jacobi_Energy_App}
\end{align}
with
\begin{align}
    \tilde{e} &= \frac{e}{h}, \\
    \tilde{i} &= \frac{i}{h}, \\
    \tilde{b} &= \frac{a-a_{\rm p}}{a_{\rm p}h}.
\end{align}
The approximation in eq.~(\ref{eq:Jacobi_Energy_App}) is valid for planetesimals close to proto-Jupiter.
At the beginning of the simulations, planetesimals have such  small eccentricities and inclinations 
that the Jacobi energies are mainly determined by semi-major axes.
The feeding zone is given only by $a$.
Therefore, $M_{\rm FZ,tot}$ is roughly given by:  
\begin{align}
    M_{\rm FZ,tot} = \int_{a_{\rm FZ,in}}^{a_{\rm FZ,out}} 2 \pi r \Sigma_{\rm solid} (t=t_0) d r,
\end{align}
where $a_{\rm FZ,in}$ and $a_{\rm FZ,out}$ are the minimum and maximum semi-major axes of the feeding zone boundary that were reached during the simulations. 
With $h$ given by the final planetary mass $M_{\rm p}= 318 M_{\oplus}$ and $a_{\rm p} = 5.2$\AU, $a_{\rm FZ,in} = (1-2\sqrt{3} h) a_{\rm p} = 3.95 \AU$ and $a_{\rm FZ,out} = (1+2\sqrt{3} h) a_{\rm p} = 6.45 \AU$ in Model-A.
Therefore, $M_{\rm FZ,tot}=46.7 M_{\oplus}$. %$M_{\rm FZ,tot}=51.7 M_{\oplus}$. 

In the No-Drag case (black solid line), 
the total mass of captured planetesimals in the simulation $M_{\rm Z,cap}$ 
exceeds $30M_{\oplus}$ and $F_{\rm cap}$ is $\sim0.7$.
Note that the fraction of the captured mass seems larger than those obtained in previous studies (i.e. $\sim0.3$ in \citet{Shibata+2019} and $\sim0.5$ in \citet{Podolak+2020} at most), however, 
here we use different capture radius (see Appendix~\ref{app:capture_radius}) and planetesimal distribution from these studies.
They substitute a larger mass than $M_{\rm FZ,tot}$ in the denominator of eq.~(\ref{eq:definition_Fcap}).
Therefore, we obtain higher $F_{\rm cap}$ than in these studies. 

Gas drag inhibits planetesimal accretion. 
The Jacobi energies of planetesimals are decreased due to  scattering by proto-Jupiter and the following eccentricity and inclination damping by gas drag, which eliminates planetesimals from the feeding zone \citep[e.g.][]{Shibata+2019}.  
Since gas drag is more profound for smaller planetesimals, $M_{\rm Z,cap}$ decreases with decreasing $R_{\rm pl}$.

Panel (a)-2 of fig.~\ref{fig:Results_psRpl} shows the fraction of captured planetesimals as a function of initial semi-major axis $f_{\rm cap}$, which is defined as: 
\begin{align}
    f_{\rm cap} = \frac{\Delta N_{\rm cap}(a_0)}{\Delta N_{0}(a_0)},
\end{align}
where $\Delta N_{\rm cap}$ is the number of captured planetesimals 
with initial semi-major axis $a_\mathrm{0}$ 
in the bin of width $\Delta a=0.05 \AU$. $\Delta N_{\rm cap}$ is the total number of planetesimals initially distributed in the bin.
The blue shaded area is the region swept by the expanding feeding zone. 
The vertical dashed lines show the centre of $j:j-1$ and $j-1:j$ mean motion resonances given as: 
\begin{align}
    a_{j:j-1} &= \left( \frac{j-1}{j} \right)^{2/3} a_\mathrm{p}, \\
    a_{j-1:j} &= \left( \frac{j}{j-1} \right)^{2/3} a_\mathrm{p}. \\
\end{align}

We find that planetesimals that are captured by proto-Jupiter mainly come from 
$a_{4:3}~<~a_0~<~a_{3:4}$, except the region around $a_{0}\sim~a_{\rm p}=5.2 \AU$ where the planetesimals are in the horseshoe orbits.
Almost no planetesimals are captured from 
$a_0<a_{4:3}$ and $a_{3:4}<a_0$, even in the feeding zone.

The relation between planetesimal accretion and the MMRs has been discussed in detail in \citet{Zhou+2007}. 
Planetesimals trapped in MMRs have regular orbital evolution and cannot easily be captured by proto-Jupiter. 
Once the overlap of adjacent MMRs occurs, the overlapping permits the chaotic orbits of planetesimals, and they can be captured. 
In $j:j-1$ MMR, a resonance overlap occurs when \citep[e.g.][]{Murray+1999}
\begin{align}
    M_{\rm p} \gtrsim \left( \frac{j}{4.7} \right)^{-7/2} M_{\rm J}. \label{eq:Mp_resonance_overlap}
\end{align}

At the begging of the simulation the resonance overlap occurs only for MMRs with $j\gtrsim11$.
As the planetary mass increases, the width of the MMRs expands and some MMRs start to overlap with adjacent MMRs. 
When the growing planet reaches Jupiter's mass, the overlapping to MMRs with $j\ga 5$ occurs.
Planetesimals located in $a_{4:3}~<~a_0~<~a_{3:4}$ are excited into the chaotic orbit by the resonance overlap and these planetesimals can be captured by proto-Jupiter.
On the other hand, planetesimals initially located in $a_0<a_{4:3}$ and $a_{3:4}<a_0$ remain trapped in the MMRs because the resonance overlap does not occur for such MMRs with small $j$. 
As a result, the planetesimal accretion efficiency  $f_{\rm cap}$ changes significantly around $a_{4:3}$ and $a_{3:4}$.
Further discussions on the effect of MMRs on the capture of planetesimals are given below.

\subsection{Planetesimal accretion onto a migrating proto-Jupiter}\label{sec:results_wMig}
In this section we present the results for Model-B.
The boundary of the feeding zone expands as the protoplanet grows and moves inward due to migration.
We find that the inner boundary of the feeding zone moves from $\sim5.81\AU$ to $3.95\AU$. 
While the outer boundary of the feeding zone first moves outward, it moves inward when $M_{\rm p} \gtrsim 40 M_{\oplus}$. 
As a result, the outer boundary of the feeding zone barely moves from the initial location of $\sim6.74\AU$.
For Model-B, we find that $a_{\rm FZ,in}=3.95 \AU$ and $a_{\rm FZ,out} = 6.80 \AU$.
The total mass of planetesimals swept by the expanding feeding zone is found to be 
$M_{\rm FZ,tot}=58.9 M_{\oplus}$.

Panel (b)-1 of fig.~\ref{fig:Results_psRpl} shows the cumulative mass of captured planetesimals as a function of time, while panel (b)-2 shows the fraction of captured planetesimals $f_{\rm cap}$ as a function of the initial semi-major axis of planetesimals. 
Comparing to the no-migration cases (Model-A), we find that $M_{\rm Z,cap}$ increases except for the case where gas drag is excluded.  

During the planetary migration planetesimals that are initially located interior to proto-Jupiter are trapped into MMRs and are shepherded by the migrating planet. 
As shown in \citet{Shibata+2020,Shibata+2022a}, the resonant trapping must be broken 
for planetesimals to be captured by the migrating planet.

The resonant trapping is broken by the resonance overlap which occurs for MMRs of 
$j\gtrsim5$ 
(eq.~(\ref{eq:Mp_resonance_overlap})).
Planetesimals that are initially located 
at $a_{4:3}~<~a_0~<~a_\mathrm{p,0}$
%between $4:3$ MMR and proto-Jupiter's orbit 
are first trapped by the MMRs of 
$j\gtrsim5$. 
As the planetary mass increases, the resonant trappings are broken by the resonance overlap and the trapped planetesimals can be captured by proto-Jupiter.
On the other hand, planetesimals that are initially located at 
 $a_0<a_{4:3}$
%interior to the $4:3$ MMR 
are stably trapped in the MMRs of $j\lesssim4$ during Jupiter's formation.
As a result, the capture fraction of planetesimals $f_{\rm cap}$ drastically changes around  $\sim5.2\AU$ where $4:3$ MMR is at the beginning of the simulations.
We find that even if proto-Jupiter migrated and the feeding zone swept a large mass of planetesimals ($M_{\rm FZ,tot}=58.9\Mear$), the captured  planetesimal mass would be similar to the case without planetary migration due to the low accretion efficiency.  
%is largely reduced from model A.
%The step-wise increase of $M_{\rm Z,cap}$ found in the panel (a) of fig.~\ref{fig:Result_wMig} rounghly correponds to the initiation of the resonance overlap.

The resonant trapping induced by planetary migration 
can be broken by the effect of overstable libration \citep{Goldreich+2014,Shibata+2020,Shibata+2022a}.
However,  overstable libration rarely occurs in our simulations since proto-Jupiter  forms almost in-situ and migrates only $\sim 1\AU$ in the radial direction. 
Overstable libration is triggered by the competition between the eccentricity damping by 
gas drag and the eccentricity excitation due to the trapped in MMRs by the migrating planet. 
The strength of the eccentricity excitation depends on the migration distance of the planet. 
If the protoplanet migrates over a large distance as in the cases presented by \citet{Shibata+2020,Shibata+2022a}, the eccentricity excitation is sufficiently strong to trigger  overstable libration.
However, in Model-B, the migration distance of proto-Jupiter is only $\sim 1\AU$, where overstable libration does not occur, and the planetesimals are stably trapped in the MMRs.

\subsection{Effect of initial eccentricities and inclinations}\label{sec:results_psEcc}
Next, we perform a parameter study where we vary the initial eccentricity and inclination of planetesimals in Model-B.

%%%%%%%%%%%%%%%%%%%%%%%%%%%%%%%%%%%%%%%%%%%%%%%%%%%%%%%%%%%%%
\begin{figure*}
  \begin{center}
    \includegraphics[width=180mm]{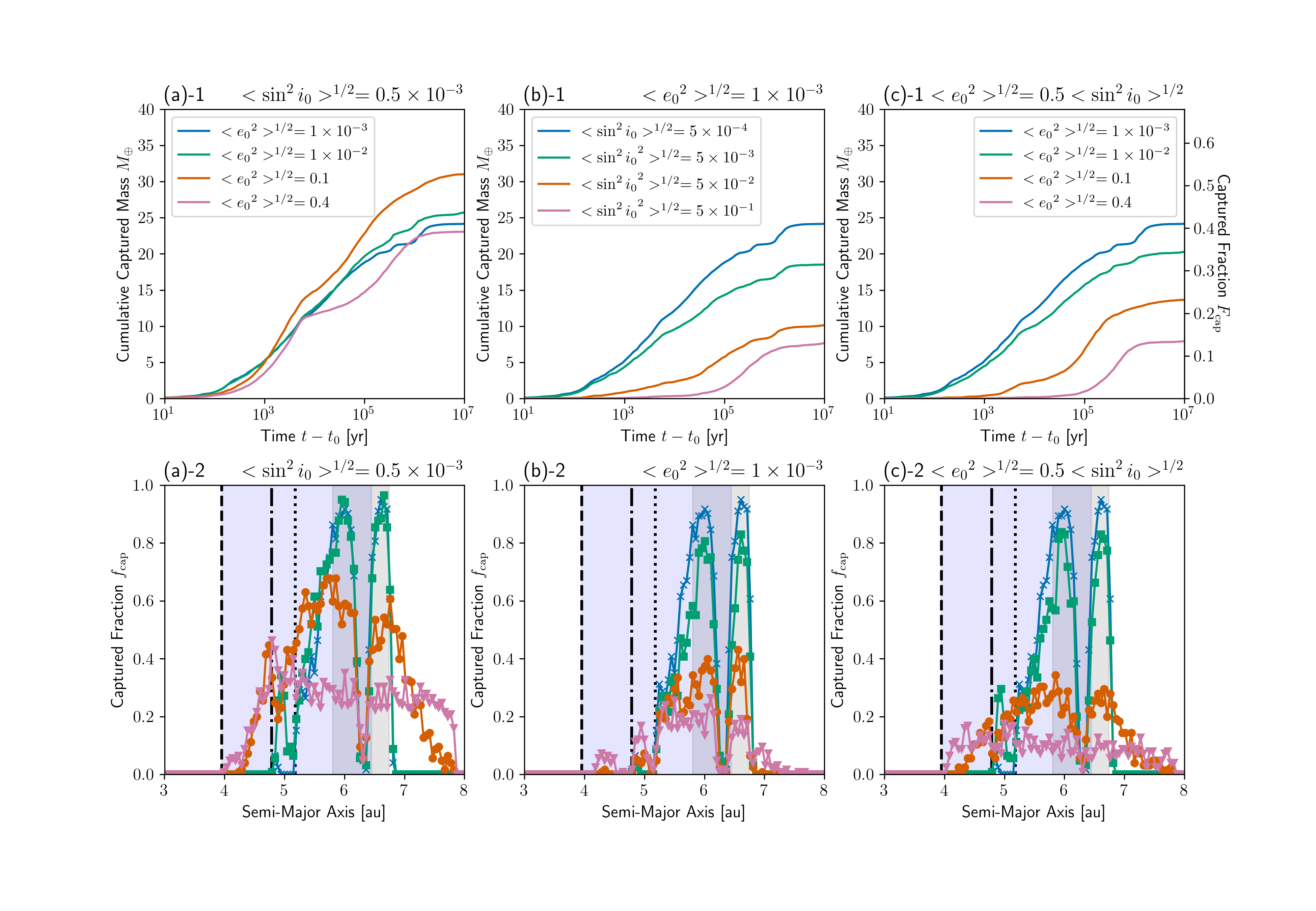}
    \caption{
    The results of the parameter study about the initial orbital profile $\langle {e_0}^2\rangle^{1/2}$ and $\langle \sin^2 i_0 \rangle^{1/2}$.
    {\bf Left column}: the results in the cases where we change $\langle {e_0}^2\rangle^{1/2}$ using the constant value of $\langle \sin^2 i_0 \rangle^{1/2}=5\times10^{-4}$.
    {\bf middle column}: the results in the cases where we change $\langle \sin^2 i_0 \rangle^{1/2}$ using the constant value of $\langle {e_0}^2\rangle^{1/2}=10^{-3}$.
    {\bf right column}: the results in the cases where we change $\langle {e_0}^2\rangle^{1/2}$ using the relation of $\langle {e_0}^2\rangle^{1/2}=2\langle \sin^2 i_0 \rangle^{1/2}$.
    Upper panels show the cumulative captured mass of planetesimals $M_\mathrm{cap}$ as a function of the calculation time $t-t_0$.
    Lower panels show the capture fraction of planetesimals $f_0$ as a function of the initial semi-major axis $a_\mathrm{pl,0}$.
    Here, we set the radius of planetesimals $R_\mathrm{pl}$ as $10^7 \cm$.
    }
    \label{fig:time_Mcap_psOrb}
  \end{center}
\end{figure*}
%%%%%%%%%%%%%%%%%%%%%%%%%%%%%%%%%%%%%%%%%%%%%%%%%%%%%%%%%%%%%

First, we change $\langle {e_0}^2\rangle^{1/2}$ from $10^{-3}$ to $0.4$ using constant $\langle \sin^2 i_0 \rangle^{1/2}=5\times10^{-4}$.
Panel (a)-1 in fig.~\ref{fig:time_Mcap_psOrb} shows the cumulative mass of captured planetesimals $M_\mathrm{cap}$ as a function of the calculation time $t-t_0$ and panel (a)-2 shows the fraction of captured planetesimals $f_0$ as a function of the initial semi-major axis $a_{\rm 0}$.
The process of planetesimal accretion can be divided into the two: the shear dominated regime where $e/h<1$ and the dispersion dominated regime where $e/h>1$.
As time progresses, the reduced hill $h$ changes from $0.02$ to $0.07$.
When $\langle {e_0}^2\rangle^{1/2}\leq10^{-2}$, planetesimal accretion occurs in the shear dominated regime and  $f_0$ is almost independent of $\langle {e_0}^2\rangle^{1/2}$.
On the other hand, when $\langle {e_0}^2\rangle^{1/2}\geq10^{-1}$, planetesimal accretion occurs in the dispersion dominated regime and $f_0$ decreases with the increasing $\langle {e_0}^2\rangle^{1/2}$.

Increasing $\langle {e_0}^2\rangle^{1/2}$ expands the region where planetesimals can be captured by proto-Jupiter.
The planetesimals captured by proto-Jupiter mainly come from the region between $5 \AU$ and $7 \AU$ in the cases of $\langle {e_0}^2\rangle^{1/2}\leq10^{-2}$, 
while planetesimals are captured by proto-Jupiter from the region between $4 \AU$ and $8 \AU$ in the cases of $\langle {e_0}^2\rangle^{1/2}\geq10^{-1}$. 
We show the 
initial and final 
feeding zone of proto-Jupiter for the planetesimals with small eccentricity ($e/h\ll1$) with gray 
($a_\mathrm{p}=a_\mathrm{p,0}$ and $M_{\rm p}=M_{\rm p,0}$) and blue ($a_\mathrm{p}=a_\mathrm{p,f}$ and $M_{\rm p}=M_{\rm p,f}$) areas, respectively.
For 
$\langle {e_0}^2\rangle^{1/2}\geq10^{-2}$, 
even planetesimals outside the indicated feeding zone can be captured because their Jacobi energy can be positive for high eccentricities as shown in eq.~(\ref{eq:Jacobi_Energy}). 
At the beginning of the simulations, the planetesimals with $\langle {e_0}^2\rangle^{1/2}\geq10^{-1}$ have larger Jacobi energy than the planetesimals with  $\langle {e_0}^2\rangle^{1/2}\leq10^{-2}$. 
Therefore, the feeding zones tend to be wider for larger $\langle {e_0}^2\rangle^{1/2}$. 

In addition, the effect of resonant shepherding is weaker for the planetesimals with higher initial eccentricity \citep[e.g.][]{Murray+1999}.
For $\langle {e_0}^2\rangle^{1/2}\leq10^{-2}$, planetesimals initially distributed 
at  $a_0<a_{4:3}$ %interior to the $4:3$ MMR of proto-Jupiter 
are trapped into the $4:3$ MMR as proto-Jupiter migrates.
As shown by eq.~(\ref{eq:Mp_resonance_overlap}), the resonance overlap does not occur and the planetesimals are stably trapped in the $4:3$ MMR during Jupiter's formation.
Therefore, planetesimals that are initially located at 
 $a_0<a_{4:3}$ %interior to the $4:3$ MMR 
are barely captured.  
For $\langle {e_0}^2\rangle^{1/2}\geq10^{-1}$, however, planetesimals in the inner disk are captured due to the insignificance of the resonant trap for high eccentricities. 
It is known that if the eccentricity of a planetesimal is higher than the critical eccentricity the planetesimal is not trapped by the MMRs. 
The critical eccentricity is given by  \citep[e.g.][]{Murray+1999}: 
\begin{align}
    e_{\rm crit} = \sqrt{6} \left[ \frac{3}{|f_{\rm d}|} \left(j-1\right)^{4/3} j^{2/3} \frac{M_{*}}{M_{\rm p}} \right]^{-1/3}. \label{eq:critical_eccentricity}
\end{align}
Here, $f_{\rm d}$ is the interaction coefficient and given as $-2.84$ for $4:3$ MMR.
For $4:3$ MMR, ${e}_{\rm crit}$ changes from 0.03 to 0.11.
Thus, planetesimals have larger eccentricities than $e_{\rm crit}$ and are captured by proto-Jupiter.
We find that as the initial $\langle {e_0}^2\rangle^{1/2}$ increases, the capture probability decreases but proto-Jupiter captures planetesimals from a wider region of the disk.
As a result, 
the total mass of captured planetesimals $M_\mathrm{cap,tot}$ takes the highest value when $\langle {e_0}^2\rangle^{1/2}=0.1$.

Second, we change $\langle \sin^2 i_0 \rangle^{1/2}$ from $5\times10^{-4}$ to $0.5$ using constant $\langle {e_0}^2\rangle^{1/2}=10^{-3}$.
Panel (b)-1 in fig.~\ref{fig:time_Mcap_psOrb} shows $M_\mathrm{cap}$ as a function of $t-t_0$ and panel (b)-2 shows $f_0$ as a function of $a_{\rm 0}$. %the cumulative mass of captured planetesimals 
$M_\mathrm{cap,tot}$ decreases with increasing $\langle \sin^2 i_0 \rangle^{1/2}$ and $f_0$ also decreases with increasing $\langle \sin^2 i_0 \rangle^{1/2}$. %The total mass of captured planetesimals decreases  the capture fraction
 
If the thickness of the planetesimal disk $a_{\rm p} \langle \sin^2 i_0 \rangle^{1/2}$ is larger than the physical radius of the growing planet $R_{\rm p}$ the capture efficiency decreases with increasing $\langle \sin^2 i_0 \rangle^{1/2}$. 
Different from the results in panel (a)-2, the region from which planetesimals are captured is rather insensitive to the initial inclination of planetesimals $\langle \sin^2 i_0 \rangle^{1/2}$.
This suggests that $\langle \sin^2 i_0 \rangle^{1/2}$ affects the capture probability, but has a negligible effect on the configuration of MMRs.

Finally, we change $\langle {e_0}^2\rangle^{1/2}$ from $10^{-3}$ to $0.4$ keeping the relation of $\langle {e_0}^2\rangle^{1/2}=2\langle \sin^2 i_0 \rangle^{1/2}$.
The equipartition of planetesimals' kinetic energy results in the relation of $\langle {e}^2\rangle^{1/2}=2\langle \sin^2 i \rangle^{1/2}$, which is easily achieved during the formation of Jupiter's core.
Panel (c)-1 of fig.~\ref{fig:time_Mcap_psOrb} shows $M_\mathrm{cap}$ as a function of $t-t_0$ and panel (c)-2 shows $f_0$ as a function of $a_{\rm 0}$. %the cumulative mass of captured planetesimals
Although the region from which planetesimals are captured expands with the increase of $\langle {e_0}^2\rangle^{1/2}$ and $\langle \sin^2 i_0 \rangle^{1/2}$, $M_{\rm Z,cap}$ decreases with $\langle {e_0}^2\rangle^{1/2}$ and $\langle \sin^2 i_0 \rangle^{1/2}$ because of the decrease in the capture probability. %the total mass of captured planetesimals 

We find that $M_\mathrm{cap,tot}$ exceeds $\sim20 M_\oplus$ with $\langle {e_0}^2\rangle^{1/2}=10^{-3}$, however, $M_\mathrm{cap,tot}$ is reduced by a factor of 2 or more if $\langle {e_0}^2\rangle^{1/2}$ is $\gtrsim10^{-1}$.
The estimated heavy-element mass in Jupiter's envelope is $\sim10 M_\oplus$ \citep[e.g.][]{Wahl+2017,Debras+2019,Stevenson+2020,Nettelmann+2021,Miguel+2022}. 
Therefore, the values of $\langle {e_0}^2\rangle^{1/2}$ and $\langle \sin^2 i_0 \rangle^{1/2}$ are quite important for the enrichment of Jupiter's envelope.
We discuss this point further in Sec.~\ref{sec:discussion_TotMass}.
}

\section{Role of mean motion resonances}\label{sec:MMRs}
{
Mean motion resonances can significantly affect the efficiency of planetesimal accretion. 
In this section, we focus on the role of mean motion resonances. 
%by analysing the results obtained in the previous section.

\subsection{Relation between capture fraction and phase angle}\label{sec:MMRs_phase}
%%%%%%%%%%%%%%%%%%%%%%%%%%%%%%%%%%%%%%%%%%%%%%%%%%%%%%%%%%%%%
\begin{figure*}
  \begin{center}
    \includegraphics[width=160mm]{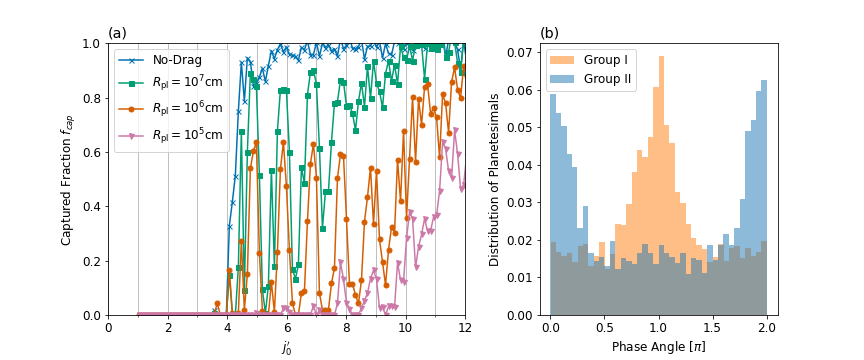}
    \caption{
    {\bf Panel (a)}:
    Fraction of captured planetesimals as a function of $j^\prime_0$ (see eq.(\ref{eq:j_definition}) for the definition) in the case of non-migrating proto-Jupiter (Model-A).
    The green, orange and magenta lines show the cases of $R_{\rm pl}=10^{7}\cm$, $10^{6}\cm$ and $10^{5}\cm$, respectively.
    The blue line shows the case where the aerodynamic gas drag is artificially neglected.
    {\bf Panel (b)}:
    Histogram of planetesimals as a function of phase angle at $t-t_0=1.6\times10^5 \yr$
    with $R_\mathrm{pl}=10^7 \cm$.
    The orange and blue lines show the planetesimals in group I (${\rm mod} (j_0^{\prime})<0.5$) and in group II (${\rm mod} (j_0^{\prime})>0.5$), respectively.
    We use 40 bins here.
    }
    \label{fig:j0_fcap_woMig}
  \end{center}
\end{figure*}
%%%%%%%%%%%%%%%%%%%%%%%%%%%%%%%%%%%%%%%%%%%%%%%%%%%%%%%%%%%%%

First, we analyse the results for Model-A. 
%for simplicity.
In the exact first order resonance ($j:j-1$ or $j-1:j$ resonance), the period ratio of planetesimals to proto-Jupiter is given by: 
\begin{align}
    \frac{P_{\rm orb}}{P_{\rm orb,p}} = \left( \frac{a}{a_\mathrm{p}} \right)^{3/2} =
        \begin{cases}
            \displaystyle{
                \frac{j-1}{j}
            } &{\rm for}~ a < a_{\rm p}, \\
            \displaystyle{    
                \frac{j}{j-1}
            } &{\rm for}~ a_{\rm p} < a,
        \end{cases}
\end{align}
where $P_{\rm orb}$ and $P_{\rm orb,p}$ are the orbital periods of the planetesimals and proto-Jupiter, respectively. 
We expand $j$ from integer to real number and define a new parameter $j^{\prime}$ as: 
\begin{align}
    %\frac{T_{\rm K}}{T_{\rm K,p}} &= \frac{j^{\prime}-1}{j^{\prime}}, \\
    j^{\prime}(a) &=
            \begin{cases}
            \displaystyle{
                \left\{ 1-\left( \frac{a}{a_{\rm p}} \right)^{3/2}\right\}^{-1}
            } &{\rm for}~ a < a_{\rm p}, \\
            \displaystyle{    
                \left\{ 1-\left( \frac{a_{\rm p}}{a} \right)^{3/2}\right\}^{-1}
            } &{\rm for}~ a_{\rm p} < a.
        \end{cases}
    \label{eq:j_definition}
\end{align}

The semi-major axes of planetesimals barely change before they are scatted by proto-Jupiter. 
If proto-Jupiter does not migrate, the resonance configurations of planetesimals relate to the initial semi-major axes or their initial $j_0^\prime = j^\prime (a_0)$. 
Panel (a) of fig.~\ref{fig:j0_fcap_woMig} shows  $f_{\rm cap}$ as a function of $j_0^\prime$.
We find a clear relationship between $f_{\rm cap}$ and the location of the MMRs.
Under the disk gas drag, the local peaks of $f_{\rm cap}$ exist around each resonance and the peaks are slightly deviated from the exact resonant centres.  
$f_{\rm cap}$ is higher in the region of $k-1/2<j^{\prime}<k$ ($k$ is the integer) than the region of $k<j^{\prime}<k+1/2$.
Hereafter, we divide planetesimals into two groups;
planetesimals with ${\rm mod} (j_0^\prime,1) < 0.5$ (group I),
and planetesimals with ${\rm mod} (j_0^\prime,1) > 0.5$ (group II), 
where the function ${\rm mod}(j_0^\prime,1)$ gives a remainder of $j_0^\prime$ divided by 1.

Around the $j:j-1$ MMR, planetesimals have a specific feature in the phase angles defined by: 
\begin{align}
    \varphi = j \lambda_{\rm p} -(j-1) \lambda -\varpi,
\end{align}
where $\lambda_{\rm p}$ and $\lambda$ are the mean longitudes of proto-Jupiter and the planetesimal, and $\varpi$ is the longitude of pericentre.
In panel (b), we show the histogram of planetesimals as a function of the phase angle $\varphi$ at $t-t_0=1.6\times10^5\yr$ for group I (orange) and  group II (blue).
We find that a large fraction of planetesimals has $\varphi\sim\pi$ in group I and $\varphi\sim0$ in group II.
Thus, fig.~\ref{fig:j0_fcap_woMig} suggests that the capture probability of planetesimals with $\varphi\sim0$ is higher than that of planetesimals with $\varphi\sim\pi$.

\subsection{Relation between phase angle and Jacobi energy}\label{sec:MMRs_Jacobi}
%%%%%%%%%%%%%%%%%%%%%%%%%%%%%%%%%%%%%%%%%%%%%%%%%%%%%%%%%%%%%
\begin{figure*}
  \begin{center}
    \includegraphics[width=160mm]{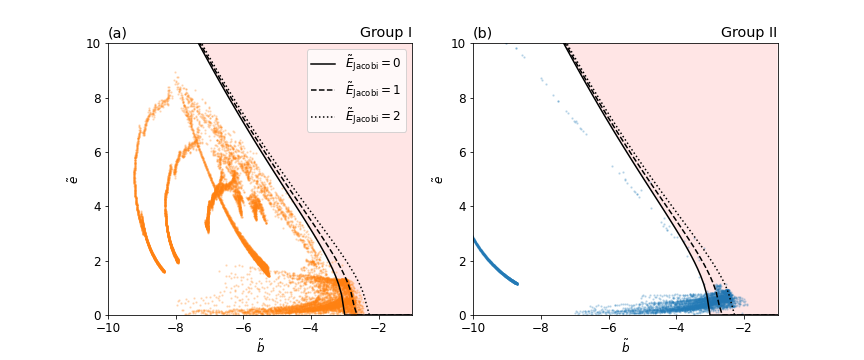}
    \caption{
    Orbital evolution of planetesimals on the plane of $\tilde{b}-\tilde{e}$.
    \textbf{Panel (a)}: the orbits of planetesimals in group I (${\rm mod}(j_0^{\prime},1)<0.5$).
    \textbf{Panel (b)}: the orbits of planetesimals in group II (${\rm mod}(j_0^{\prime},1)>0.5$).
    The red shaded area is the feeding zone with $\tilde{E}_{\rm Jacobi}>0$ and the black solid line shows the boundary of the feeding zone ($\tilde{E}_{\rm Jacobi}=0$).
    The dashed and dotted lines are $\tilde{E}_{\rm Jacobi}=1$ and $2$, respectively.
    We show the orbits of 15 planetesimals in the case of Model-A with $R_{\rm pl}=10^7 \cm$.
    }
    \label{fig:b_eh}
  \end{center}
\end{figure*}
%%%%%%%%%%%%%%%%%%%%%%%%%%%%%%%%%%%%%%%%%%%%%%%%%%%%%%%%%%%%%
\begin{figure*}
  \begin{center}
    \includegraphics[width=160mm]{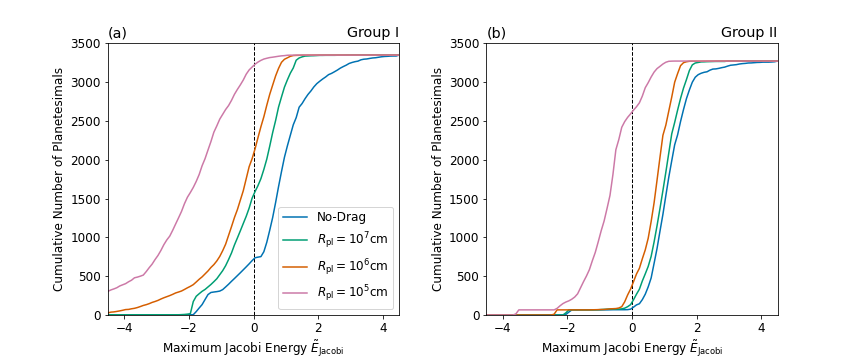}
    \caption{
    The cumulative number of planetesimals as a function of the maximum Jacobi energy achieved in the simulations.
    Here, we show the results in Model-A.
    \textbf{Panel (a)}: the planetesimals in group I (${\rm mod}(j_0^{\prime},1)<0.5$).
    \textbf{Panel (b)}: the planetesimals in group II (${\rm mod}(j_0^{\prime},1)>0.5$).
    }
    \label{fig:Ejmax}
  \end{center}
\end{figure*}
%%%%%%%%%%%%%%%%%%%%%%%%%%%%%%%%%%%%%%%%%%%%%%%%%%%%%%%%%%%%%

We next investigate the connection between the phase angle $\varphi$ and the capture fraction $f_0$.
Figure \ref{fig:b_eh} shows the orbital evolution of 15 planetesimals that are initially distributed interior to proto-Jupiter's orbit.
We plot $\tilde{b}$ and $\tilde{e}$ %for $x$ and $y$ axis. 
for planetesimals in group I (${\rm mod} (j_0^\prime,1) < 0.5$) in panel (a) and for those in group II (${\rm mod} (j_0^\prime,1) > 0.5$) in panel (b).
The growth of proto-Jupiter induces the decrease of $|\tilde{b}|$ or the increase of the Jacobi energy. 
Thus, planetesimals approach proto-Jupiter from left to right in fig.~\ref{fig:b_eh}. 
Planetesimals are in the planetary feeding zone, if the Jacobi energie becomes positive.
However, strong scatterings by  proto-Jupiter pump up their eccentricities, which induce effective gas drag. 
The eccentricity damping by gas drag then  reduces their Jacobi energies. 
Therefore, strong scatterings mainly remove  planetesimals from the feeding zone due to the decrease of their Jacobi energies.  

We find that planetesimals in Group II (or planetesimals with $\varphi\sim0$) enter deeper into the feeding zone with 
high Jacobi energies as $\tilde{E}_{\rm Jacobi}\sim2$ (dotted line), 
while planetesimals in Group I (or planetesimals with $\varphi\sim\pi$) are scattered by proto-Jupiter before they  reach the deep feeding zone. 
It is known that a resonance of a planetesimal in MMR with $\varphi=0$ is more stable to perturbations, such as gas drag, than that with $\varphi=\pi$ \citep[e.g.][]{Murray+1999}.
Due to the stable resonant trapping, planetesimals with $\varphi\sim0$ keep their eccentricity small 
without scatterings 
and enter deeper regions in the feeding zone until the resonance overlap destabilizes their orbits.

In order to investigate this effect further, we focus on the maximum Jacobi energy achieved in the simulations $\tilde{E}_{\rm Jacobi,max}$.
Figure \ref{fig:Ejmax} shows the cumulative number of planetesimals as a function of $\tilde{E}_{\rm Jacobi,max}$ of planetesimals. 
Planetesimals in group I ($\varphi\sim\pi$) are shown in the left panel and those in group II ($\varphi\sim0$) are shown in the right panel.
Here, we present only planetesimals that are initially outside the feeding zone. 
The cumulative number of planetesimals at $E_\mathrm{Jacobi,max} = 0$ indicates the number of planetesimals that did not reach the feeding zone.
In the case of group I ($\varphi\sim\pi$), many planetesimals could not enter the feeding zone. 
This is because those planetesimals are scattered into the eccentric orbit before they can enter the feeding zone as shown in fig.~\ref{fig:b_eh}.
On the other hand, for group II ($\varphi\sim0$) almost all the planetesimals enter the feeding zone except the case of $R_{\rm pl}=10^5 \cm$.
Even in the case of $R_{\rm pl}=10^5 \cm$, the planetesimals can have higher Jacobi energies than the planetesimals in group I.
Figure \ref{fig:Ejmax} shows that the MMRs control the inflow flux of planetesimals into the feeding zone, and that more planetesimals enter the feeding zone if the resonance angle librates around $\varphi\sim0$.

We find that $\tilde{E}_{\rm Jacobi,max}$ tends to be smaller for smaller planetesimals because of gas drag.
The planetesimal accretion rate is given by the product of the surface density of planetesimals inside the feeding zone and the capture probability \citep[e.g.][]{Chambers2006}.
For small planetesimals, eccentricities and inclinations are damped by gas drag, so that high capture probabilities are expected \citep[e.g.][]{Inaba+2001}.
However, small planetesimals tend to stay outside the feeding zone as discussed above, leading to a rather low surface density inside the feeding zone. 
As a result, the accretion rate is smaller for the smaller planetesimals.

\subsection{The case of planetary migration}\label{sec:MMR_inMigration}
%%%%%%%%%%%%%%%%%%%%%%%%%%%%%%%%%%%%%%%%%%%%%%%%%%%%%%%%%%%%%
\begin{figure}
  \begin{center}
    \includegraphics[width=80mm]{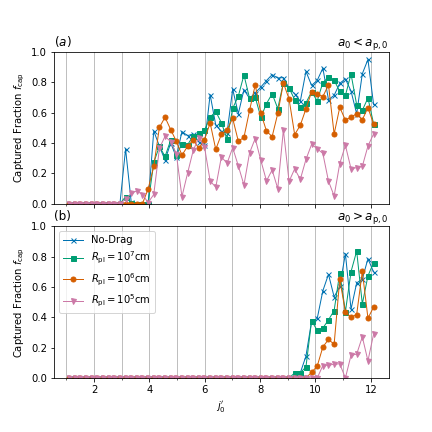}
    \caption{
    Same as panel (a) in fig.~\ref{fig:j0_fcap_woMig}, but for Model-B.
    We show planetesimals which initially locate interior to proto-Jupiter's orbit in panel (a), 
    and those exterior to proto-Jupiter's orbit in panel (b).
    }
    \label{fig:j0_fcap_wMig}
  \end{center}
\end{figure}
%%%%%%%%%%%%%%%%%%%%%%%%%%%%%%%%%%%%%%%%%%%%%%%%%%%%%%%%%%%%%

Here we analyse the results for Model-B where proto-Jupiter slightly migrates inward during the gas accretion phase.
In this case, the planetesimals which are initially located interior to proto-Jupiter's orbit are trapped by MMRs and the resonance angles converge into $\varphi=0$.
On the other hand, planetesimals which initially locate exterior to proto-Jupiter's orbit are not trapped by MMRs and the resonance angles do not converge.

Figure~\ref{fig:j0_fcap_wMig} shows $f_{\rm cap}$ as a function of $j_0^{\prime}$. 
Here we plot $f_{\rm cap}$ for the planetesimals of $a_{\rm pl,0}< a_{\rm p,0}$ (upper panel) and $a_{\rm pl,0}> a_{\rm p,0}$ (lower panel), respectively.
In the upper panel, the jagged profiles of $f_{\rm cap}$ found in the no-migration cases (see fig.~\ref{fig:j0_fcap_woMig}) are smoothed.
This is because almost all planetesimals have the same resonance angle ($\varphi\sim0$).
On the other hand, planetesimals that are initially located exterior to proto-Jupiter are not trapped by the MMRs. 
In this case only the planetesimals initially located in the feeding zone ($j_0^\prime\gtrsim10$) can be captured by proto-Jupiter.
}

\subsection{Mean motion resonances in excited planetesimal disk}\label{sec:MMR_inExcitedDisk}
%%%%%%%%%%%%%%%%%%%%%%%%%%%%%%%%%%%%%%%%%%%%%%%%%%%%%%%%%%%%%
\begin{figure}
  \begin{center}
    \includegraphics[width=80mm]{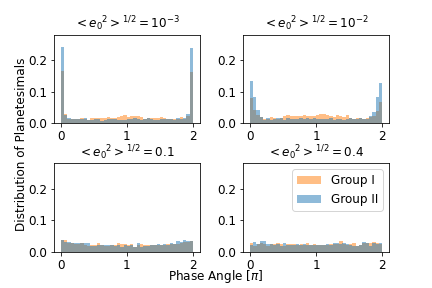}
    \caption{
    Same as panel (b) in fig.~\ref{fig:j0_fcap_woMig}, but show the cases in Model-B and of $\langle {e_0}^2 \rangle^{1/2}=2\langle {\sin }^2 i_0 \rangle^{1/2}=10^{-3}$, $10^{-2}$, $0.1$ and $0.4$.
    }
    \label{fig:phase_angle_psEcc}
  \end{center}
\end{figure}
%%%%%%%%%%%%%%%%%%%%%%%%%%%%%%%%%%%%%%%%%%%%%%%%%%%%%%%%%%%%%

The role of mean motion resonances in planetesimal accretion has been investigated by various groups.  %\citep{Zhou+2007, Shiraishi+2008, Shibata+2020, Shibata+2022a}.
\citet{Zhou+2007} suggested that mean motion resonances affect the gas accretion rate during  phase 2 since the  planetesimal accretion rate changes frequently due to the resonance overlap.
\citet{Shibata+2020,Shibata+2022a} showed that shepherding of planetesimals by the mean motion resonances is important for predicting the metallicity of hot/warm-Jupiters. 
Note, however, that the effect of mean motion resonances is weak when $\langle {e_0}^2 \rangle$ is large as shown in fig.~\ref{fig:time_Mcap_psOrb}.

Figure~\ref{fig:phase_angle_psEcc} shows the distribution of phase angles in the cases of $\langle {e_0}^2 \rangle^{1/2}=2\langle {\sin }^2 i_0 \rangle^{1/2}=10^{-3}$, $10^{-2}$, $0.1$ and $0.4$ in Model-B.
When $\langle {e_0}^2\rangle^{1/2}\leq10^{-2}$, there is a peak around $\phi\sim0$ and these planetesimals are trapped in MMRs.
However, the peak is almost flattened when $\langle {e_0}^2\rangle^{1/2}\geq10^{-1}$.
The planetesimals are not in resonant trapping and MMRs hardly affect  planetesimal accretion. 
This is because eccentricities of planetesimals are higher than the critical eccentricity $e_\mathrm{crit}$ which changes from 0.03 to 0.11 during Jupiter's formation.
Thus, we find that the effect of the MMRs is important when $\langle {e_0}^2\rangle^{1/2} < e_\mathrm{crit}$, but negligible when $\langle {e_0}^2\rangle^{1/2}\gtrsim e_\mathrm{crit}$.
Also, note that the role of mean motion resonances investigated in previous studies does not apply for a  disk with excited planetesimals.

\section{Comparison with the analytical expressions }\label{sec:comparison}
{
A semi-analytical approach of the planetesimal accretion is used for modelling the formation of the heavy-element core \citep{Pollack+1996,Alibert+2005,Fortier+2013} and the envelope enrichment of gas giant planets \citep{Shiraishi+2008,Hasegawa+2018,Hasegawa+2019,Venturini+2020}.
Here, we compare our numerical results with various published semi-analytical accretion rates. 
Further details on the semi-analytical approach can be found in the appendix \ref{app:analytical}.

%%%%%%%%%%%%%%%%%%%%%%%%%%%%%%%%%%%%%%%%%%%%%%%%%%%%%%%%%%%%%
\begin{figure*}
  \begin{center}
    \includegraphics[width=160mm]{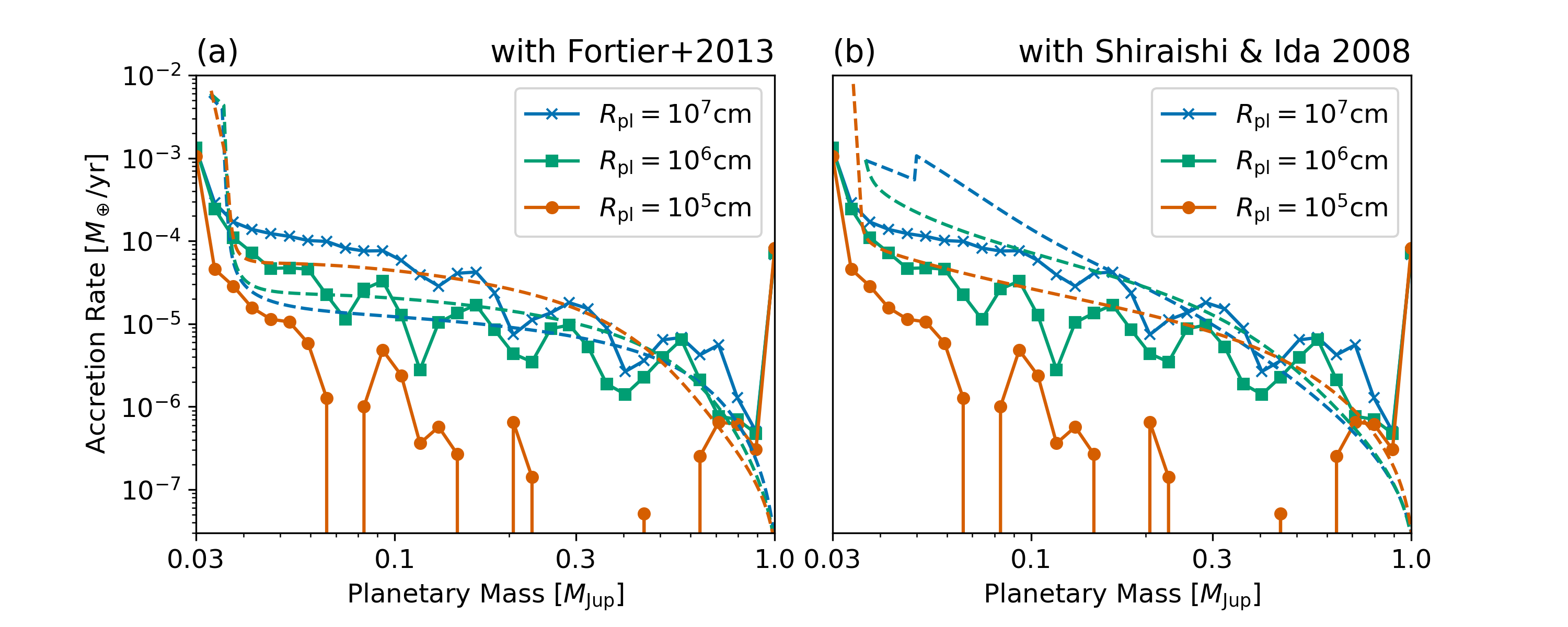}
    \caption{
    Planetesimal accretion rate as a function of the planetary mass in Model-A.
    In both panels, the results  of this study are plotted with solid lines.
    {\bf Panel (a)}: the accretion rate obtained by the semi-analytical formulae in \citet{Fortier+2013} is plotted with dashed lines.
    {\bf Panel (b)}: the accretion rate obtained by the semi-analytical formulae in \citet{Shiraishi+2008} is plotted with dashed lines.
    }
    \label{fig:AccRate}
  \end{center}
\end{figure*}
%%%%%%%%%%%%%%%%%%%%%%%%%%%%%%%%%%%%%%%%%%%%%%%%%%%%%%%%%%%%%
The planetesimal accretion rate of a protoplanet is given by: 
\begin{align}\label{eq:Analitycal_Collision_Rate}
    \frac{d  M_{\rm cap}}{d t} \propto \Sigma_{\rm solid} P_{\rm col},
\end{align}
where $P_{\rm col}$ is the non-dimensional collision probability.
Using a statistical approach \citep[e.g.][]{Inaba+2001}, $P_{\rm col}$ can be  estimated as a function of $\tilde{e}$, $\tilde{i}$ and $\tilde{r}=R_{\rm cap}/R_{\rm H}$ where $R_{\rm H}$ is the planetary hill radius. 
Following the evolution of $\tilde{e}$, $\tilde{i}$ and $\tilde{r}$, \citet{Fortier+2013} obtains the planetesimal accretion rate starting from a small  core up to a fully-formed giant planet.
In their model, the surface density of planetesimals around the protoplanet $\Sigma_{\rm solid}$ is estimated in a simple manner where the effect of the disk gas drag is neglected (see the appendix for  details).

In panel (a) of fig.~\ref{fig:AccRate}, we show the planetesimal accretion rates from the N-body simulations (solid line) and the semi-analytical model by \citet{Fortier+2013} (dashed line).
Here, we show the cases where proto-Jupiter does not migrate  (Model-A).
In the statistical model, the planetesimal accretion rate is higher for the smaller planetesimals.
Interestingly, this trend on $R_{\rm pl}$ is opposite to the results from the N-body simulations.
Smaller planetesimals experience stronger drag from the disk gas and the equilibrium $\tilde{e}$ and $\tilde{i}$ also decrease. 
The collision probability $P_{\rm col}$ is higher for the smaller $\tilde{e}$ and $\tilde{i}$, and therefore the accretion rate is higher for the smaller planetesimals in the statistical model.
Even in the N-body simulations, $P_{\rm col}$ is larger for  smaller $\tilde{e}$ and $\tilde{i}$, which is supported by the results in Sec.~\ref{sec:results_psEcc}.
However, for smaller planetesimals, it is more difficult to enter the deeper feeding zone as shown in sec.~\ref{sec:MMRs_Jacobi}.
This means that $\Sigma_{\rm solid}$ would be smaller for the smaller planetesimals, but this effect is not included in the statistical approach.

To account for the effect of gas drag on $\Sigma_{\rm solid}$, \citet{Shiraishi+2008} derived a semi-analytical expression for the planetesimal accretion rate by fitting the results of N-body simulations.
In their model, $\Sigma_{\rm solid}$ is scaled by the growth timescale of the protoplanet and the damping timescale from disk gas. 

Panel (b) of fig.~\ref{fig:AccRate} shows the planetesimal accretion rate inferred by the N-body simulations (solid line) and the semi-analytical model of  \citet{Shiraishi+2008} (dashed line).
Here, we show the results for the cases where proto-Jupiter does not migrate (Model-A).
Unlike in panel (a), here the planetesimal accretion rate is higher for the larger planetesimals, which is consistent to the results of the N-body simulations.

%%%%%%%%%%%%%%%%%%%%%%%%%%%%%%%%%%%%%%%%%%%%%%%%%%%%%%%%%%%%%
\begin{figure*}
  \begin{center}
    \includegraphics[width=160mm]{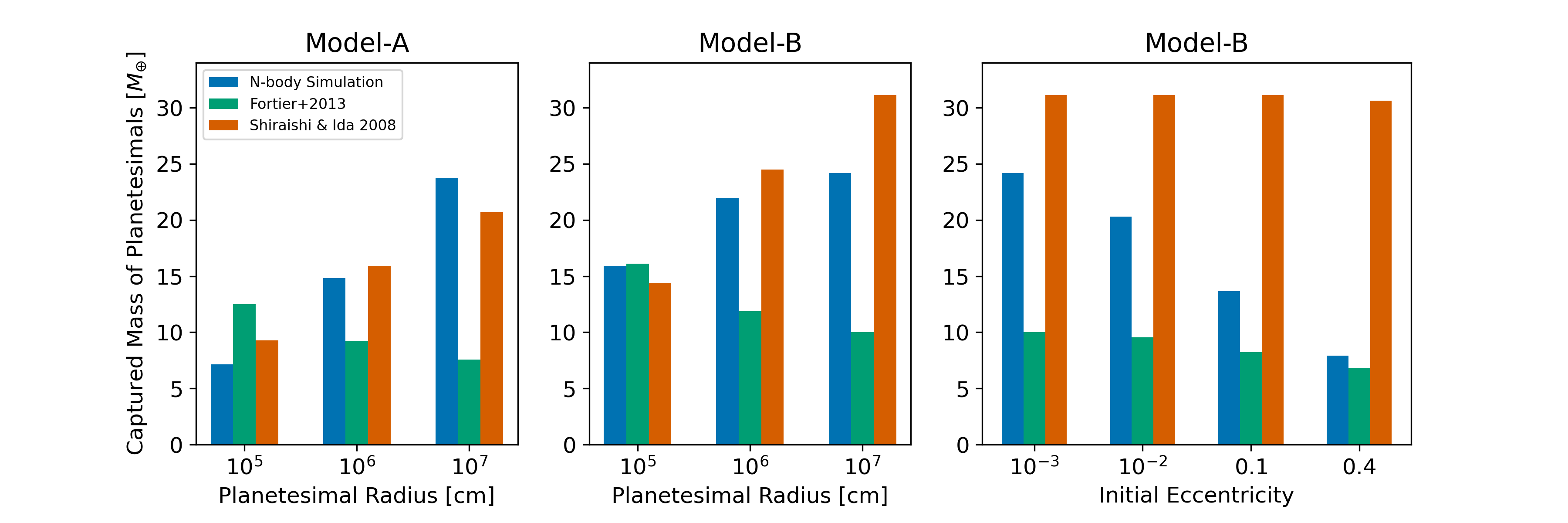}
    \caption{
    A comparison of the total mass of captured planetesimals obtained by N-body simulations and semi-analytical models.
    The blue, green, and orange bars correspond to the results obtained by the N-body simulations, the semi-analytical model of \citet{Fortier+2013}, and that of \citet{Shiraishi+2008}, respectively. 
    The left and middle panels show the results for Model-A and Model-B, respectively, as a function of the planetesimal size.
    The right panel shows the results of the parameter study in regard to $<e_{0}^2>^{1/2}$ with the relation of  $<e_{0}^2>^{1/2}=2<\sin^2 i_{0}>^{1/2}$.
    }
    \label{fig:Result_Mtot_Compare}
  \end{center}
\end{figure*}
%%%%%%%%%%%%%%%%%%%%%%%%%%%%%%%%%%%%%%%%%%%%%%%%%%%%%%%%%%%%%
Figure \ref{fig:Result_Mtot_Compare} shows the total mass of captured planetesimals $M_\mathrm{cap,tot}$ in each model.
In the left and middle panels, we compare the results in Model-A and Model-B, respectively.
While there are differences of several Earth-masses, the semi-analytical model by \citet{Shiraishi+2008} reproduces the dependency on the planetesimals size found in N-body simulations.
However, as found in the right panel, the total mass of captured planetesimals is almost independent of the value of $<e_{0}^2>^{1/2}$ in the semi-analytical models.
In the semi-analytical model, planetesimals entering the feeding zone are assumed to suffer from the strong gravitational scattering from the protoplanet.
In this case, $\tilde{e}$ and $\tilde{i}$ exceeds $\sim1$ quickly regardless of the initial values $<e_{0}^2>^{1/2}$ and $<\sin^2 i_{0}>^{1/2}$.
This assumption is not always true as shown in fig.~\ref{fig:b_eh}, where planetesimals enter the feeding zone before being scattered by the protoplanet.
$\tilde{e}$ and $\tilde{i}$ of the planetesimals entering the feeding zone would be determined by the MMRs rather than the strong gravitational scattering from the protoplanet.

We find much incompleteness in the semi-analytical models.
We conclude that the development of improved new semi-analytical models is required. 
%is quite important, but it is out of the scope of this study. We will work on it in the future work.
}

\section{Discussion}\label{sec:discussion}
{

\subsection{Total heavy-element mass}\label{sec:discussion_TotMass}

\subsubsection{Mass of captured planetesimals in the runaway gas accretion phase}
%%%%%%%%%%%%%%%%%%%%%%%%%%%%%%%%%%%%%%%%%%%%%%%%%%%%%%%%%%%%%
\begin{figure}
  \begin{center}
    \includegraphics[width=80mm]{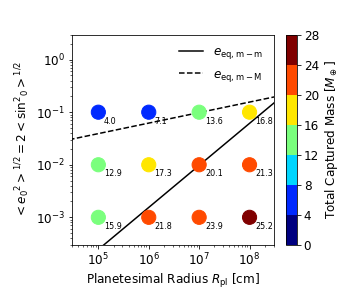}
    \caption{
    Total mass of captured planetesimal $M_\mathrm{cap,tot}$ as a function of the planetesimals size $R_{\rm pl}$, and the initial eccentricity of planetesimals $\langle {e_0}^2\rangle^{1/2}$.
    The colour code corresponds to $M_\mathrm{cap,tot}$ of each simulation according to the right colour bar.
    We also plot the value of $M_\mathrm{cap,tot}$ in the lower-right side of each plot.
    We show the results in the cases with $\langle {e_0}^2\rangle^{1/2}=2\langle \sin^2 i_0 \rangle^{1/2}$.
    The solid and dashed lines show the equilibrium eccentricity of planetesimals without embryos $e_{\rm eq,m-m}$ and with embryos $e_{\rm eq,M-m}$, respectively.% given by eq.~(\ref{eq:ecc_ec_small}).
    }
    \label{fig:Mtot}
  \end{center}
\end{figure}
%%%%%%%%%%%%%%%%%%%%%%%%%%%%%%%%%%%%%%%%%%%%%%%%%%%%%%%%%%%%%

As we show in Sec.~\ref{sec:planetesimal_accretion}, the planetesimal accretion rate depends on $R_{\rm pl}$, $\langle {e_0}^2\rangle^{1/2}$, and $\langle \sin^2 i_0 \rangle^{1/2}$.
Assuming that the energy equipartition $\langle {e_0}^2\rangle^{1/2}=2\langle \sin^2 i_0 \rangle^{1/2}$ is achieved in the planetesimal disk before runaway gas accretion, we perform additional simulations where we change $R_{\rm pl}$ and $\langle {e_0}^2\rangle^{1/2}=2\langle \sin^2 i_0 \rangle^{1/2}$.
Figure~\ref{fig:Mtot} shows the total mass of captured planetesimals $M_{\rm cap,tot}$. 
As we find in Sec.~\ref{sec:planetesimal_accretion}, $M_{\rm cap,tot}$ is larger for the larger $R_{\rm pl}$ and smaller $\langle {e_0}^2\rangle^{1/2}=2\langle \sin^2 i_0 \rangle^{1/2}$.

In the single-sized planetesimal disk, the eccentricity and inclination of the planetesimals are determined by their mutual scattering and the damping of disk gas drag.
Equating the stirring timescale and the damping timescale, \citet{Fortier+2013} found that the equilibrium eccentricity of planetesimals $e_{\rm eq,m-m}$ is give by: 
\begin{align}\label{eq:ecc_ec_small}
    e_{\rm eq,m-m} = 2.31 \left( \frac{{m}_{\rm pl}^{4/3} \Sigma_{\rm solid} a {\rho_{\rm pl}}^{2/3}}{C_{\rm d} \rho_{\rm gas} {M_{*}}^{2}} \right)^{1/5},
\end{align}
where $m_{\rm pl}$ is the mass of planetesimals, 
$\rho_{\rm pl}$ is the density of planetesimals, 
$\rho_{\rm gas}$ is the density of disk gas, and
$C_\mathrm{d}$ is the non-dimensional gas drag coefficient.
The gravitational scattering from embryos also excites the eccentricity of planetesimals.
Around embryos with mass of $M_{\rm emb}$, the equilibrium eccentricity $e_{\rm eq,m-M}$ is given by:  \citep{Thommes+2003}
\begin{align}
    e_{\rm eq,m-M} = 1.7 \left( \frac{m_{\rm pl}^{1/3} {\rho_{\rm pl}}^{2/3}}{b C_{\rm d} \rho_{\rm gas} a} \right)^{1/5} \left( \frac{M_{\rm emb}}{M_{*}} \right)^{1/3}, \label{eq:ecc_ec_mM}
\end{align}
where $b$ is the full width of the feeding zone.
If embryos are formed and distributed in the region from $4\AU$ to $8\AU$ in the oligarchic regime, $\langle {e_0}^2\rangle^{1/2}$  would be as large as $e_{\rm eq,m-M}$.
On the other hand, if embryos form only around Jupiter's core, $\langle {e_0}^2\rangle^{1/2}$ would be smaller in the region farther from Jupiter's core.
Thus, the value of $\langle {e_0}^2\rangle^{1/2}$ would be between the values of $e_{\rm eq,m-M}$ and $e_{\rm eq,m-m}$.

In Fig.~\ref{fig:Mtot} we present $e_{\rm eq,m-m}$ with solid line and $e_{\rm eq,m-M}$ with dashed line.
Here we use $\Sigma_{\rm solid}=20 \g/\cm^2$, $a=5.2\AU$, $\rho_{\rm gas}=1\times10^{-11} \g/\cm^3$, $C_{\rm d}=2$, $b=10$, and $M_{\rm emb}=1M_{\oplus}$.
In previous studies \citep{Zhou+2007,Shiraishi+2008,Shibata+2019,Podolak+2020}, $\langle {e_0}^2\rangle^{1/2}$ and $\langle \sin^2 i_0 \rangle^{1/2}$ are set as small values as $\sim10^{-3}$.
The total mass of captured planetesimals $M_\mathrm{cap,tot}$ would be $16$-$25 M_\oplus$ in this case.
If $\langle {e_0}^2\rangle^{1/2}$ is increased by the scattering of embryos and given by eq.~(\ref{eq:ecc_ec_mM}), $M_\mathrm{cap,tot}$ is reduced by more than a factor of two and $M_{\rm cap,tot}$ could be significantly smaller. 
The formation of embryos in addition to Jupiter's core is expected to occur in a massive planetesimal disk  as considered here.
Then the mass of accreted planetesimals during runaway gas accretion  would be significantly smaller than estimated  previously  due to the higher eccentricities and inclinations of planetesimals.

\subsubsection{Cases of a massive core formation}\label{sec:discussion_Mp0}
In the simulations presented above, we set the initial mass of proto-Jupiter $M_\mathrm{p,0}$ as $10 M_\oplus$,  assuming that at this mass runaway gas accretion begins.
However, the onset of runaway gas accretion could occur at  higher masses  \citep[e.g.,][]{Movshovitz+2010, Lozovsky+2017}.
We therefore perform additional simulations changing $M_\mathrm{p,0}$ and assuming $\langle {e_0}^2\rangle^{1/2}=2\langle \sin^2 i_0 \rangle^{1/2}=e_{\rm eq,m-M}$.
Figure~\ref{fig:Mtot} shows the total mass of captured planetesimals $M_\mathrm{cap,tot}$. 
By setting the mass of Jupiter's core to be $M_\mathrm{core,0}=0.5 M_\mathrm{p,0}$, we also show the total heavy-element mass in Jupiter $M_\mathrm{Z,Jup} = M_\mathrm{core,0} + M_\mathrm{cap,tot}$ with the solid line.

$M_\mathrm{cap,tot}$ is smaller for larger $M_\mathrm{p,0}$ because more planetesimals inside the initial feeding zone are depleted by the formation of Jupiter's core. 
However, this effect is cancelled by the massive core formation and $M_\mathrm{Z,Jup}$ increases with $M_\mathrm{p,0}$. %the total mass of heavy elements in Jupiter 
We find that $M_\mathrm{Z,Jup}$ changes with $M_\mathrm{p,0}$ despite the fixed total heavy-element mass (core + planetesimals) in our disk model.

Due to the high eccentricity of planetesimals, a long time is required  to deplete the planetesimals inside the feeding zone by accretion. 
We define the depletion timescale of planetesimals due to accretion by proto-Jupiter  $\tau_\mathrm{cap}$ by:
\begin{align}
    \tau_\mathrm{cap} = \frac{M_\mathrm{FZ,tot}}{\dot{M}_\mathrm{cap}}.
\end{align}
$\tau_\mathrm{cap}$ can be longer than $10^{5} \yr$ when $\langle {e_0}^2\rangle^{1/2}=2\langle \sin^2 i_0 \rangle^{1/2}$ is as large as $0.1$.
Once proto-Jupiter enters runaway gas accretion, the gas accretion timescale    $\tau_\mathrm{acc}$ becomes shorter than $\tau_\mathrm{cap}$ due to the rapid accretion of gas. 
The feeding zone expands prior to the effective capture of planetesimals by  proto-Jupiter, and planetesimals that are not captured would be scattered to eccentric orbits and be eliminated from the feeding zone.
We find that the fraction of captured planetesimals $f_\mathrm{cap}$ is  $\sim 20\%$ (see fig.~\ref{fig:Results_psMp0} in Appendix~\ref{sec:results_MassiveCore}).
On the other hand, $\tau_\mathrm{acc}$ could be longer than $\tau_\mathrm{cap}$ until the onset of runaway gas accretion as shown in \citet{Movshovitz+2010}.
Large fraction of planetesimals would be depleted prior to the expansion of the feeding zone.
Our results suggest that the accretion efficiency of heavy elements (core+planetesimals) depends on the timing of the onset of runaway gas accretion.
}

\subsubsection{Comparison with the interior models of Jupiter}

\begin{figure}
  \begin{center}
    \includegraphics[width=80mm]{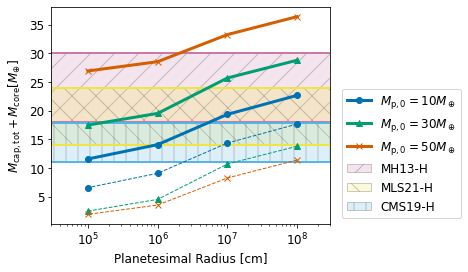}
    \caption{
    Total mass of heavy elements estimated from our N-body simulations $M_\mathrm{Z,Jup}=M_\mathrm{cap,tot}+M_\mathrm{core,0}$ as a function of the planetesimals size $R_{\rm pl}$.
    The blue, green, and orange solid lines show the cases of $M_\mathrm{p,0}=10 M_\oplus$, $30 M_\oplus$, and $50 M_\oplus$, respectively.
    We also plot the total mass of captured planetesimals $M_\mathrm{cap,tot}$ with the dashed lines.
    The red, yellow and blue shaded areas correspond to the bulk heavy-element mass estimated with the measured gravitational moments by the Juno spacecraft with different equations of state \citep{Miguel+2022}.
    }
    \label{fig:Mtot_psM0}
  \end{center}
\end{figure}

The measured gravitational moments by the Juno spacecraft can be used to constrain Jupiter's interior structure and for estimating the total heavy-element mass $M_\mathrm{Z,Jup}$  \citep[e.g.][]{Wahl+2017,Debras+2019,Stevenson+2020,Nettelmann+2021,Miguel+2022}.
\citet{Miguel+2022} found that $M_\mathrm{Z,Jup}$ is $\sim11$-$30\Mear$ and explored the sensitivity of the inferred heavy-element mass to the used equation of state (EOS).
In \citet{Miguel+2022}, three EOSs were considered:   \citet{Militzer+2013} (MH13-H), \citet{Chabrier+2019} (CMS19-H) and \citet{Mazevet+2020} (MLS21-H). 
$M_\mathrm{Z,Jup}$ is estimated as $18$-$30~M_\oplus$ with MH13-H, $14$-$24M_\oplus$ with MLS21-H and $11$-$18~M_\oplus$ with CMS19-H.
Fig.~\ref{fig:Mtot_psM0} shows the estimated $M_\mathrm{Z,Jup}$ with the shaded areas.

Our results are consistent with $M_\mathrm{Z,Jup}$ for the large parameter space used in this study.
For the estimated heavy-element mass inferred by the MH13-H EOS, a core larger than $15 M_\oplus$, or planetesimals with $R_\mathrm{pl}\gtrsim10^{7}$ are required.
If hydrogen is indeed denser in Jupiter's interior conditions as suggested by the MLS21-H and CMS19-H EOSs or even denser as implied by Quantum Monte Carlo simulations \citep{2018PhRvL.120b5701M}, it would suggest that Jupiter's core mass is smaller than $15 M_\oplus$. 
It is clear that an improved  understanding of the hydrogen (and hydrogen-helium) EOS could  further constrain Jupiter's heavy-element mass, and therefore   its origin  \citep[e.g.,][]{2020NatRP...2..562H}.

\subsection{Perturbations on mean motion resonances}\label{sec:MMRs_embryos}

In our simulations, we change the initial eccentricity $\langle {e_0}^2\rangle^{1/2}$ and inclination $\langle \sin^2 {i_0}\rangle^{1/2}$ to account for the effect of embryos' scattering.
However, the scattering from embryos not only increases $\langle {e_0}^2\rangle^{1/2}$ and $\langle \sin^2 {i_0}\rangle^{1/2}$, but also adds perturbations on resonance angles.
The perturbations on resonance angles accelerate the break of resonant trapping \citep{Malhotra1993b,Tanaka+1999,Shibata+2022a}.
If we include the embryos' scattering in N-body simulations directly, the effects of MMRs might be further diminished.

In sec.~\ref{sec:discussion_TotMass}, we set $\langle {e_0}^2\rangle^{1/2}=e_\mathrm{eq,m-M}$, which is comparable to or lager than $e_\mathrm{crit}$ in our disk model.
In these cases, almost all planetesimals are not in resonant trapping (see fig.~\ref{fig:phase_angle_psRpl}).
Thus, the effects of MMRs are negligibly small, and we expect that the planetesimal accretion rate would not change from our results even if we include the embryos' scattering directly.
If proto-Jupiter grows almost solely as considered in previous studies, $\langle {e_0}^2\rangle^{1/2}$ might be smaller than $e_\mathrm{crit}$ and MMRs play an important role on planetesimal accretion.
Such formation scenario would be possible if Jupiter's core migrated from outer disk region \citep{Bosman+2019,Oberg+2019,Shibata+2022}.

In addition to the embryos' scattering, the collisions of planetesimals also add perturbations on resonance angles and break resonant trapping \citep{Malhotra1993b}.
Also, the collisions might trigger the break-up of planetesimals and generate smaller fragments.
This effect is known to accelerate planetesimal accretion rate in the oligarchic regime because collision probability $P_\mathrm{col}$ increases with decreasing $\langle {e_0}^2\rangle^{1/2}$.
However, planetesimal accretion rate decreases with $R_\mathrm{pl}$ in the runaway gas accretion phase 
because smaller planetesimals are found to be easily eliminated from the feeding zone.
Thus, whether the planetesimal collisions increase or decrease planetesimal accretion rate is unclear and must be investigated in future work.

\subsection{Giant impacts of embryos}\label{sec:Discussion_GI}
The existence of other embryos reduces the available heavy-element mass that can be added to the planet via planetesimal accretion. Note, however, that the mass of heavy elements in the form of embryos is significantly smaller (\citet{Kobayashi+2021}) than in the form of planetesimals. %$\sim 10\%$ 
Nevertheless, giant impacts of planetary embryos during runaway gas accretion could increase the planetary metallicity \citep[e.g.,][]{Ginzburg+2020,Ogihara+2021}.
\citet{Liu+2019} found that proto-Jupiter can capture $\sim 40 \%$ of the embryos as it grows. 
%However, it was assumed that the embryos are in co-planar orbits while in reality the orbits of  embryos should be inclined which reduces the impact probability.
As a result, giant impacts of embryos could add a few Earth-masses of heavy elements to the growing Jupiter.
Also, \citet{Liu+2019} suggested that Jupiter's fuzzy core and bulk composition could be a result of a giant impact where the impactor's mass is $10 M_\oplus$. 
Other embryos that are formed in the disk are expected to have masses of a few Earth-masses  \citep{Kobayashi+2021}.
These embryos can growth further in mass via planetesimal accretion. However, it remains unclear whether such massive embryos can indeed form.  This topic should be investigated in detail in future work.

Embryos more massive than $1 M_{\oplus}$ migrate faster than proto-Jupiter \citep[e.g.][]{Kanagawa+2018}.
In this case, unlike in the case of planetesimals, the orbit between the embryos and proto-Jupiter converges if the embryo's orbit is exterior to the orbit of proto-Jupiter.
The embryo can then be trapped by MMRs, which would increase the capture efficiency inferred in sec.~\ref{sec:MMRs_Jacobi}.
The dynamical friction from the planetesimals would also enhance the probability to capture embryos due to the reduction of the embryos' eccentricities. 
Thus, the interaction between planetesimals and embryos also affects the predicted enrichment of Jupiter's envelope.
We hope to investigate this topic in detail in future research.

\subsection{In-situ formation vs.~outer-disk formation}\label{sec:Discussion_VS}

\begin{figure}
  \begin{center}
    \includegraphics[width=80mm]{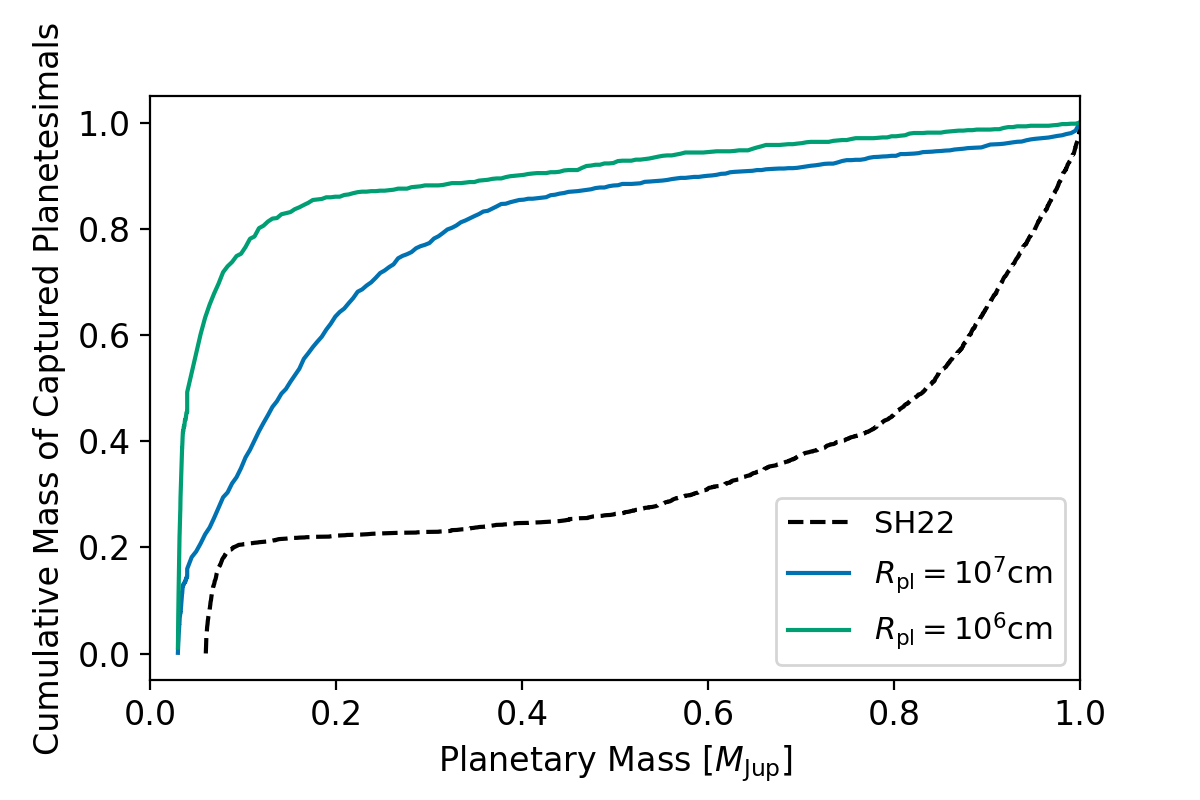}
    \caption{
    Cumulative mass of captured planetesimals $M_\mathrm{cap}$ normalized by the total mass of captured planetesimals $M_\mathrm{cap,tot}$ as a function of the planetary mass $M_{\rm p}$.
    The result obtained by \citet{Shibata+2022} is shown with the black dashed line.
    The blue and green solid lines show the results for $R_{\rm pl}=10^7 \cm$ and $R_{\rm pl}=10^6~\cm$, respectively.
    The cases presented correspond to $\langle {e_0}^2\rangle^{1/2}=2\langle \sin^2 {i_0} \rangle^{1/2}=~e_{\rm eq, M-m}$ and $M_\mathrm{p,0}=10 M_\oplus$.
    SH22 started their simulation with $M_\mathrm{p,0}=20 M_\oplus$
    }
    \label{fig:Discussion_dotM}
  \end{center}
\end{figure}

{
In this study, we assume that Jupiter had formed nearly in-situ, which is supported by studies showing that 
the gas accretion timescale is shorter than the migration timescale during Jupiter's formation \citep{Tanaka+2020}. 
However, currently there is no way to discriminate among different formation locations for Jupiter.  
Several models suggested that Jupiter formed much farther from its current location. 
While the accretion efficiency is low \citep{Okamura+2021}, Jupiter's core might be formed in outer disk region via pebble accretion. 
For example, for a scenario where Jupiter's core is formed via pebble accretion it was shown that with a massive supply of pebbles ($\gtrsim100M_{\oplus}$ in total), the core can reach pebble isolation mass, which exceeds $20 M_{\oplus}$ if the disk's aspect ratio is higher than $\sim0.05$, in several tens $\AU$ within 2 Myr \citep{Lambrechts+2014,Bitsch+2015,Bitsch+2019a}. 
In the following planetary migration phase, some of such planetary cores can reach current Jupiter's orbit before the disk dissipation.

The difference between the migration distance between the case of in-situ formation and formation in the outer disk comes from the different gas accretion and planetary migration rates. 
In \citet{Bitsch+2015,Bitsch+2019a}, the maximum gas accretion rate is limited to $80\%$ of the disk accretion rate because the disk gas can cross the gap opened by the planet \citep{Lubow+2006} and the planetary migration timescale is scaled by the disk's viscous timescale.
However, a slower migration model due to the gap opening is adopted in \citet{Tanaka+2020}.
The difference of the fraction of the migration timescale to the gas accretion timescale makes the difference in the migration distance between the models of \citet{Tanaka+2020} and  \citet{Bitsch+2015,Bitsch+2019a}.
It is therefore required to better determine the gas accretion and the planetary migration rates and their associated physics. 

The mixing of heavy elements in the planetary envelope could be a  tracer of Jupiter's formation history. 
Since recent interior models of Jupiter imply that the planet is not fully convective \citep[e.g.,][]{Leconte+2013, Wahl+2017,Vazan+2018,Debras+2019}, the heavy-element distribution  today might reflect the planetesimal accretion rate during its  formation. 
If Jupiter formed much farther away, migration from the outer disk to its current orbit is expected to trigger a late planetesimal bombardment \citep[][hereafter SH22]{Shibata+2022}.
In this formation model  proto-Jupiter was assumed to form at $\sim$ 20\AU~and to migrate to its current location after the onset of  runaway gas accretion. 
This formation pathway leads to planetesimal accretion during the planetary migration.   
Therefore, many planetesimals are expected to be deposited in the outer envelope rather than in the deep interior.
This late accretion phase can provide several M$_{\oplus}$ of planetesimals even when the surface density of planetesimals is smaller than $1 \g/\cm^2$ around $10 \AU$.

We compare our results with the result of SH22.
The planetesimal disk model used in SH22 is different from the disk model used here. 
In order to compare the results from these two studies, we plot the cumulative mass of captured planetesimals normalized by the total mass of captured planetesimals $\tilde{M}_\mathrm{cap}=M_\mathrm{cap}/M_\mathrm{cap,tot}$ in fig.~\ref{fig:Discussion_dotM}.
The figure shows the obtained $\tilde{M}_\mathrm{cap}$ as a function of planetary mass $M_\mathrm{p}$.
We clearly see that planetesimals accreted at smaller (larger) $M_\mathrm{p}$ are deposited in deeper (shallower) regions of the envelope. 
We show the cases where $\langle {e_0}^2\rangle^{1/2}=2\langle \sin^2 {i_0} \rangle^{1/2}=e_{\rm eq, M-m}$ and $M_\mathrm{p,0}=10 M_\oplus$.
Interestingly, we identify a clear difference in $\tilde{M}_\mathrm{cap}$ between the cases of in-situ formation and formation in the outer disk followed by migration. 
In this study,  $\tilde{M}_\mathrm{cap}$ increases rapidly when $M_{\rm p}$ is small while $\tilde{M}_\mathrm{cap}$ mainly increases after a large amount of gas has accumulated in SH22.
This means that the accreted  planetesimals are deposited in the deeper envelope in the {\it in-situ} formation case, while in the case of formation in the outer disk, the accerted material is deposited in the upper envelope (atmosphere). 

As a result, we suggest that  determining  the heavy-element distribution in Jupiter's envelope could be used to determine its formation location. 
Unfortunately, structure models are non-unique and in addition, linking the current-state structure of Jupiter with its origin is challenging since the heavy-element distribution can change as the planet evolved \citep[e.g.,][]{Vazan+2018,Muller+2020a,Helled+2022}. 
A better understanding of heavy-element accretion and convective mixing during  Jupiter's evolution could therefore reveal important information on its formation history.

}

\section{Summery and conclusions}\label{sec:Conclusion}
{
We investigate planetesimal accretion onto proto-Jupiter in the  massive planetesimal disk scenario. 
Our model includes the distribution of planetesimals obtained by \citet{Kobayashi+2021}, where the surface density of solid materials exceeds $20 \g/\cm^2$ around $6\AU$.  % and a planetary core of $10 M_{\oplus}$ forms within $2\times10^5 \yr$,
We consider two formation scenarios: 
Model-A where proto-Jupiter does not migrate during the simulations and Model-B where proto-Jupiter slightly migrates inward from $6\AU$ to $5\AU$. 
%{\bf here talk about Model-A and Model-B... formation model where proto-Jupiter slightly migrates inward from $6\AU$ to $5\AU$,}
We next investigate the effect of the initial eccentricity $\langle {e_0}^2\rangle^{1/2}$ and inclination $\langle \sin^2 i_0 \rangle^{1/2}$ of planetesimals on the accretion rate.
We focus on the role of MMRs which have significant effects on the planetesimal accretion rate. 
Finally, we compare our N-body simulations with commonly used semi-analytical models for calculating the planetesimal accretion rate.

Our main conclusions are summarised as follows:
(1) Proto-Jupiter can accrete tens M$_{\oplus}$ of heavy elements by the end of runaway gas accretion in a massive planetesimal disk.
(2) The increase of $\langle {e_0}^2\rangle^{1/2}$ weakens resonant trapping leading to an enhancement of planetesimal accretion.
However, the increase of $\langle {e_0}^2\rangle^{1/2}$ and $\langle \sin^2 i_0 \rangle^{1/2}$ reduces the capture probability at the same time.
As a result, the captured mass of planetesimals decreases with the increase of $\langle {e_0}^2\rangle^{1/2}$ and $\langle \sin^2 i_0 \rangle^{1/2}$.
(3) The efficiency of planetesimal accretion changes with the resonance configuration. 
More planetesimals are captured for the planetesimals whose resonance angles librate around $\varphi\sim0$ in comparison to the planetesimals with $\varphi\sim\pi$.
The resonance angles of planetesimals are largely affected by the planetary migration.
When the orbits of planetesimals and proto-Jupiter converge, the resonance angles become $\varphi\sim0$ and the capture fraction increases.
(4) Existing semi-analytical models cannot reproduce the results obtained by N-body simulations.
The effect of the disk gas drag must be considered in order to properly estimate the surface density of planetesimals. %and their  collision probability. 
Also, the eccentricity and inclination of planetesimals entering the planetary feeding zone cannot be reproduced in the semi-analytical approach.

In the massive planetesimal disk scenario, planetary embryos are expected to form. 
If many embryos are formed around proto-Jupiter and are in oligarchic regime, the total mass of captured planetesimals $M_\mathrm{cap,tot}$ is obtained as $2-18 M_\oplus$, 
which is smaller than that obtained with the low eccentricities and inclinations assumed in previous studies. 
Assuming that the total heavy elements mass in Jupiter $M_\mathrm{Z,Jup}$ is the summation of the initial core mass $M_\mathrm{core,0}$ and the total mass of captured planetesimals $M_\mathrm{cap,tot}$, we estimate $M_\mathrm{Z,Jup}$ and find that $M_\mathrm{Z,Jup}$ increases with the initial core mass $M_\mathrm{core,0}$ even in the same planetesimal disk.
We compare our results with $M_\mathrm{Z,Jup}$ inferred from Jupiter's interior structure models.  
The inferred $M_\mathrm{Z,Jup}$ is consistent with our numerical results over a wide parameter region.
Further determination of Jupiter's internal structure could further  constrain Jupiter's formation history and put some limits on its primordial  core mass and the dominated size of accreted planetesimals. 
Finally, when comparing our in-situ formation model with a model where proto-Jupiter formed in the outer disk region, we find that the formation history does not only affect Jupiter's bulk metallicity, but also the expected heavy-element distribution in Jupiter's envelope. 

Our results demonstrate the importance of the embryos' scattering and the initial core mass of Jupiter.
Despite the low capture probability of planetesimals due to the excited eccentricities and inclinations, we find that Jupiter can accrete much heavy elements by the formation of a massive planetesimal disk.
While Jupiter's bulk metallicity can, in principle, be reproduced in the in-situ formation scenario, it remains unclear whether this model can explain the heavy-element distribution in the planet. 
We suggest that evolution models should follow the heavy-element distribution in Jupiter's envelope in order to assess whether this scenario is realistic, and we hope to address it in future research.
Finally, we suggest that information on Jupiter's  primordial core mass and composition gradients can reveal critical information on Jupiter's formation history \citep{Helled+2022}. 
}

\section*{Acknowledgements}

We thank the anonymous reviewer reading and constructive comments. 
SS and RH acknowledge support from the Swiss National Science Foundation (SNSF) under grant $200020\_188460$.
HK was supported by JSPS KAKENHI Grant Numbers 22H00179, 22H01278, 21K03642, 20H04612, 18H05436 and 18H05438.
Numerical computations were carried out on the Cray XC50 at the Center for Computational Astrophysics, National Astronomical Observatory of Japan.

\section*{Data Availability}
The data obtained in our simulations can be made available on reasonable request to the corresponding author.

%\vspace{5mm}
%\facilities{HST(STIS), Swift(XRT and UVOT), AAVSO, CTIO:1.3m,

%\software{astropy \citep{2013A&A...558A..33A,2018AJ....156..123A},  
%          Cloudy \citep{2013RMxAA..49..137F}, 
%          Source Extractor \citep{1996A&AS..117..393B}
%          }

\appendix
\section{Details of our numerical model}\label{app:method}
\subsection{Equation of motion}
In N-body simulations, we follow the orbital evolution of planetesimals.
The equation of motion is given by: 
\begin{align}
    \frac{{\rm d^2} {\bf r}_{i} }{{\rm d} t^2} &=   - \mathcal{G} \frac{M_{*}}{{r_{i{\rm,s}}}^3} {\bf r}_{i{\rm,s}} 
                                                    - \mathcal{G} \frac{M_{\rm p}}{{r_{i{\rm,p}}}^3} {\bf r}_{i{\rm,p}} 
                                                    -\frac{{\bf u}}{\tau_{\rm aero}},
\end{align}
where $\mathbf{r}_{i,\mathrm{p}}$ and $\mathbf{r}_{i,\mathrm{s}}$ are the position vector of the particle $i$ relative to the planet and central star, respectively,
$M_{*}$ and $M_{\rm p}$ are the mass of the central star and a protoplanet, respectively,
$\mathcal{G}$ is the gravitational constant, and 
$\tau_{\rm aero}$ is the gas drag timescale given by: 
\begin{align}
    \tau_{\rm aero} &=  \frac{2 m_{\rm pl}}{ C_{\rm d} \pi R_{\rm pl}^2 \rho_{\rm gas} u}. \label{eq:aerodynamic_gas_drag_timescale}
\end{align}
Here ${\bf u}={\bf v}_{\rm pl}-{\bf v}_{\rm gas}$ ($u = |{\bf u}|$) is the planetesimal's velocity (${\bf v}_{\rm pl}$) relative to the ambient gas (${\bf v}_{\rm gas}$), %$\tau_{\rm aero}$ is the timescale of aerodynamic gas drag, 
$m_{\rm pl}$ is the planetesimal's mass, 
$C_{\rm d}$ is the non-dimensional drag coefficient, 
$\rho_{\rm gas}$ is the gas density, and 
$R_{\rm pl}$ is the planetesimal's radius. 
$C_{\rm d}$ can be represented by a constant for planetesimal-sized bodies. 
However, it has different values for supersonic headwinds, low gas densities, and so on.
In order to cover all gas drag regimes, we use an approximated formula for $C_{\rm d}$ introduced by \citet{Tanigawa+2014}. % (see Appendix).

An approximated formula for $C_{\rm d}$ is given by: \citep[e.g.][]{Tanigawa+2014} % the non-dimensional drag coefficient 
\begin{align}\label{eq:Drag_Coefficient_Tanigawa}
	C_{\rm d} \simeq \left[ \left( \frac{24}{\mathcal{R}} + \frac{40}{10+\mathcal{R}} \right)^{-1} + \frac{3 \mathcal{M}}{8} \right]^{-1} + \frac{ (2-\omega) \mathcal{M}}{1+\mathcal{M}} + \omega,
\end{align}
where $\mathcal{R}$ is the Reynolds number, $\mathcal{M}$ is the Mach number, and $\omega$ is a correction factor.
These parameters are given by: 
\begin{align}
    \mathcal{R} &= \frac{2 R_{\rm pl} u}{\nu_{\rm l}},\\
    \mathcal{M} &= \frac{u}{c_{\rm s}},\\
    \omega &= 
        \begin{cases}
                \displaystyle{ 0.4
                } &{\rm for}~\mathcal{R} < 2 \times 10^5, \\
                \displaystyle{ 0.2           
                } &{\rm for}~\mathcal{R} > 2 \times 10^5,
        \end{cases}
\end{align}
where $\nu_{\rm l}$ is the kinetic viscosity.
For the ideal gas, $\nu_{\rm l}$ is obtained by: 
\begin{align}
    \nu_{\rm l} &= \frac{1}{3} \sqrt{\frac{8}{\pi}} c_{\rm s} l_{\rm p},% \\
%        &= \frac{1}{3} \frac{m_{\rm mol}}{\sigma_{\rm mol}} c_{\rm s},
\end{align}
where $l_{\rm p}$ is the mean free path of a molecule in the disk gas.
Here, we use $l_{\rm p}=m_{\rm mol}/\sigma_{\rm mol} \rho_{\rm gas}$ where $m_{\rm mol}$ and $\sigma_{\rm mol}$ are the mass and collision cross-section of a molecule, and we use the values of hydrogen molecules.

To determine $\rho_{\rm gas}$ for the gas drag, 
we assume the vertically isothermal disk and $\rho_{\rm gas}$ is given as
\begin{align}
    \rho_{\rm gas} = \frac{\Sigma_{\rm gas}}{\sqrt{2 \pi} h_{\rm s} } \exp \left( -\frac{z^2}{2 {h_{\rm s}}^2} \right),
\end{align}
where $\Sigma_{\rm gas}$ and $h_{\rm s}$ are the surface density profile of disk gas and the disk gas scale height, respectively. 
According to the force balance with the density profile, 
the disk gas rotates with the sub-Kepler velocity which is  given by: 
\begin{align}
    v_{\rm gas} = v_{\rm K} \left(1 - \eta \right), 
\end{align}
with
\begin{align}
    \eta &= \frac{1}{2} \left( \frac{h_{\rm s} }{r} \right)^2 \left[ \frac{3}{2} \left( 1 - \frac{z^2}{{h_{\rm s}}^2} \right) +\alpha +\beta \left( 1 +\frac{z^2}{{h_{\rm s}}^2} \right) \right], \label{eq:eta_disk} \\
    \alpha &= - \frac{{\rm d} \ln \Sigma_{\rm gas}}{{\rm d} \ln r}, \\
    \beta &= - \frac{{\rm d} \ln c_{\rm s}}{{\rm d} \ln r},
\end{align}
where $r$ is the radial distance from the central star and $c_{\rm s}$ is the disk's sound speed. 
In eq.~(\ref{eq:eta_disk}), we neglect $(z/h_{\rm s})^4$ and higher order terms, because gas drag is less significant at high altitudes. 
The velocity and density of the ambient disk gas are calculated from the protoplanetary disk model (see sec.~\ref{sec:method_disk}).

\subsection{Gaseous disk model}\label{app:method_gas_disk}
%\subsubsection{Gas disk}
Our disk model is based on the self-similar solution for the surface density profile of a gaseous disk  \citep{Lynden-Bell+1974}.
The mid-plane temperature of disk gas $T_{\rm disk}$ is given by: 
\begin{align}
    T_{\rm disk} = T_{\rm disk,0} \left( \frac{r}{1 \AU} \right)^{-2\beta}, \label{eq:temperature_profile}
\end{align}
where we set $T_{\rm disk,0}=200 \K$ and $\beta=1/4$.
We adopt the $\alpha$-viscosity model of \citet{Shakura+1973} where the disk gas viscosity $\nu$ is: 
\begin{align}
    \nu = \alpha_{\rm acc} c_{\rm s} h_{\rm s}. \label{eq:viscosity}
\end{align}
In this case, the disk gas viscosity $\nu=\alpha_{\rm acc} c_{\rm s} h_{\rm s}$ is proportional to $r$ and the self-similar solution $\Sigma_{\rm SS}$ is given by: 
\begin{align}
    \Sigma_{\rm SS} = \frac{M_{\rm tot,0}}{2 \pi {R_{\rm d}}^{2}} \left( \frac{r}{R_{\rm d}} \right)^{-1} T^{-3/2} \exp \left( -\frac{r}{T R_{\rm d}} \right), \label{eq:self-similar-solution}
\end{align}
with
\begin{align}
    T &= 1 + \frac{t}{\tau_{\rm vis}}, \label{eq:normalized_time} \\
    \tau_{\rm vis} &= \frac{{R_{\rm d}}^2}{\nu_{\rm d}}, \label{eq:viscous_characteristic_timescale}
\end{align}
where $M_{\rm tot,0}$ is the disk's total mass at $t=0$, $R_{\rm d}$ is a radial scaling length of protoplanetary disk, $\tau_{\rm vis}$ is the characteristic viscous timescale, and $\nu_{\rm d}$ is the disk gas viscosity at $r=R_{\rm d}$.
The surface density profile of gaseous disk is altered by the gap opening around the planet, the gas accretion onto the planet, and the disk depletion. 
We include these effects and calculate the surface density profile of disk gas $\Sigma_{\rm gas}$ by: 
\begin{align}
    \Sigma_{\rm gas} = f_{\rm gap} f_{\rm acc} f_{\rm dep} \Sigma_{\rm SS}, \label{eq:Sigma_gas}
\end{align}
where $f_{\rm gap}$ is the gap opening factor, $f_{\rm acc}$ is the gas accretion factor, and $f_{\rm dep}$ is the disk depletion factor.
For the gap opening factor, we adopt the empirically obtained model by \citet{Kanagawa+2017}.
The gap structure changes with the radial distance from the planet $\Delta r=|r-r_{\rm p}|/r_{\rm p}$ and $f_{\rm gap}$ is written as a function of $\Delta r$ by: 
\begin{align}
    f_{\rm gap} =
        \begin{cases}
            \displaystyle{            
            \frac{1}{1+0.04K} 
            } &{\rm for}~ \Delta r < \Delta R_1, \\
            \displaystyle{            
            4.0 {K^{\prime} }^{-1/4} \Delta r -0.32
            } &{\rm for}~\Delta R_1 < \Delta r < \Delta R_2 , \\
            \displaystyle{            
            1
            } &{\rm for}~ \Delta R_2 < \Delta r,
        \end{cases} \label{eq:gap_structure}.
\end{align}
with
\begin{align}
    K           &= \left( \frac{M_{\rm p}}{M_{*}} \right)^2 \left( \frac{h_{\rm p}}{r_{\rm p}} \right)^{-5} {\alpha_{\rm acc}}^{-1}, \\
    K^{\prime}  &= \left( \frac{M_{\rm p}}{M_{*}} \right)^2 \left( \frac{h_{\rm p}}{r_{\rm p}} \right)^{-3} {\alpha_{\rm acc}}^{-1}, \\
    \Delta R_1  &= \left\{ \frac{1}{4(1+0.04K)} +0.08 \right\} {K^{\prime}}^{1/4}, \\
    \Delta R_2  &= 0.33 {K^{\prime}}^{1/4}.
\end{align}
In the disk region inner to the planet, the disk's surface density is reduced by the gas accretion onto the planet.
When the gas accretion rate is given by eq.~(\ref{eq:GasAccretionRate_TT}) and the gap structure is given by eq.~(\ref{eq:gap_structure}), $f_{\rm acc}$ is written as \citep{Tanaka+2020}: 
\begin{align}
    f_{\rm acc} = 
        \begin{cases}
            1 &{\rm for}~r > r_{\rm p}, \\
            \displaystyle{            
            \left\{ 1 + \frac{D}{3\pi \nu (1+0.04K)} \right\}^{-1}
            } &{\rm for}~r \leq r_{\rm p}.
        \end{cases} \label{eq:sweet_spot}
\end{align}
To account for disk depletion processes, such as photo-evaporation or disk wind, we set the disk depletion factor $f_{\rm dep}$ to: 
\begin{align}
    f_{\rm dep} = \exp \left( - \frac{t}{\tau_{\rm dep}} \right).
\end{align}
The surface density of disk gas at the bottom of the gap, which is used for the gas accretion rate and migration rate, is obtained by  $\Sigma_{\rm gap}=\Sigma_{\rm gas} (r=r_{\rm p})$ using eq.~(\ref{eq:Sigma_gas}).

\begin{figure*}
  \begin{center}
    \includegraphics[width=180mm]{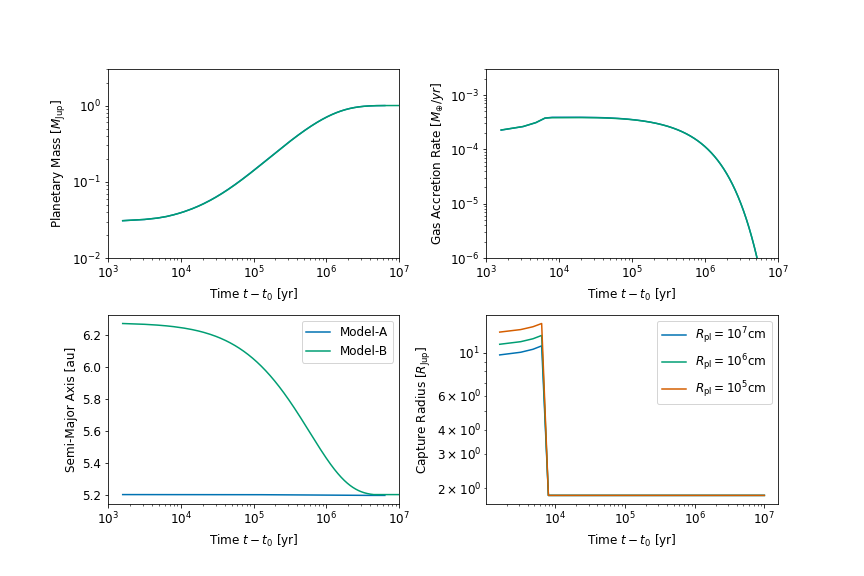}
    \caption{
    Formation model of Jupiter used in this study.
    {\bf (a):} Mass of proto-Jupiter,
    {\bf (b):} gas accretion rate onto proto-Jupiter,
    {\bf (c):} semi-major axis of proto-Jupiter, and
    {\bf (d):} capture radius of proto-Jupiter
    as a function of calculation time $t-t_0$.
    }
    \label{fig:formation_model}
  \end{center}
\end{figure*}

Using the disk model presented above, we can simulate Jupiter's growth as shown in fig.~\ref{fig:formation_model}.

%%%%%%%%%%%%%%%%%%%%%%%%%%%%%%%%%%%%%%%%%%%%%%%%%%%%%%%%%%%%%
%%%%%%%%%%%%%%%%%%%%%%%%%%%%%%%%%%%%%%%%%%%%%%%%%%%%%%%%%%%%%

\subsection{Planetesimal disk model}\label{app:method_planetesimal_disk}
The surface density profile of planetesimals  $\Sigma_\mathrm{solid}$ we use in this study is given by: 
\begin{align}
    \Sigma_\mathrm{solid} = \Sigma_\mathrm{KT21,solid} - \Sigma_\mathrm{core}, \label{eq:sigma_solid}
\end{align}
with, 
\begin{align}
    \Sigma_\mathrm{core} =
            \begin{cases}
            \displaystyle{            
            0
            } &{\rm for}~ r < a_{\rm FZ,in}, a_\mathrm{FZ,out} < r, \\
            \displaystyle{            
            \frac{0.5 M_\mathrm{p,0}}{S_\mathrm{FZ}}
            } &{\rm for}~a_{\rm FZ,in} < r < a_{\rm FZ,out}, \\
        \end{cases} \label{eq:sigma_core}
\end{align}
where $\Sigma_\mathrm{KT21,solid}$ is the solid surface density obtained in \citet{Kobayashi+2021}, and $S_{\rm FZ}$ is the area of the feeding zone.
The values of $a_{\rm FZ,in}$, $a_{\rm FZ,out}$, and $S_\mathrm{FZ}$ are calculated at $t=t_0$.

The super-particles are distributed in a given radial region, where the inner and outer edges are denoted by $a_{\rm pl,in}$ and $a_{\rm pl,out}$, respectively.
We set $a_{\rm pl,in}$ and $a_{\rm pl,out}$ as super-particles cover the region of planetary feeding zone during the simulations; namely:
\begin{align}
    a_{\rm pl,in}  &= a_{\rm p,f} \left( 1-2\sqrt{3} \left( \frac{M_{\rm p,f}}{3M_{*}} \right)^{1/3} \right), \label{eq:inner_edge_of_planetesimal_disk} \\
    a_{\rm pl,out} &= a_{\rm p,0} \left( 1+2\sqrt{3} \left( \frac{M_{\rm p,f}}{3M_{*}} \right)^{1/3} \right), \label{eq:outer_edge_of_planetesimal_disk}
\end{align}
where $a_{\rm p,0}$ and $a_{\rm p,f}$ are the initial and final semi-major axis of the protoplanet and $M_{\rm p,f}$ is the final planet's mass. 
The surface number density of super-particles $n_{\rm s}$ is given by: % $\propto r^{-3/2}$. 
\begin{align}
    n_{\rm s} = n_{\rm s,0} \left( \frac{r}{1 \AU} \right)^{-\alpha_{\rm sp}}
\end{align}
with
\begin{align}
    n_{\rm s,0} = \frac{N_{\rm sp}}{2 \pi} \frac{2 -\alpha_{\rm sp}}
                  {\left(a_{\rm pl,out}/1\AU\right)^{2-\alpha_{\rm sp}} -\left(a_{\rm pl,in}/1\AU\right)^{2-\alpha_{\rm sp}} } \left[\frac{1}{\AU^2}\right], 
\end{align}
where $N_{\rm sp}$ is the number of super-particles used in a given simulation and $\alpha_{\rm sp}=1$, with the  super-particles being distributed uniformly in the radial direction. 
We set $N_{\rm sp}=9600$ where the spatial density is kept larger than $2,000$ super-particles per $1\AU$. 
The mass per super-particle $M_{\rm sp}$ is given by: 
\begin{align}
    M_{\rm sp} (a_{\rm 0}) = \frac{\Sigma_{\rm solid} (a_{\rm 0})}{n_{\rm s} (a_{\rm 0})} ,
\end{align}
where $a_{\rm 0}$ is the initial semi-major axis of the super-particle.

\subsection{Capture radius of proto-Jupiter}\label{app:capture_radius}
During the orbital integration, we judge that a super-particle has been captured by the planet once (i) the super-particle enters the planet's envelope or (ii) its Jacobi energy (see Eq.~(\ref{eq:Jacobi_Energy}) for the definition) becomes negative in the Hill sphere.
The planetary envelope expands during runaway gas accretion. 
To account for the effect of the  envelope expansion on the planetesimal accretion, we use the approximation for the capture radius $R_{\rm cap}$ inferred by  \citep[e.g.][]{Valletta+2021}.
Fig.~\ref{fig:formation_model} shows the evolution of $R_{\rm cap}$ in our model.

\section{Semi-analytical approach}\label{app:analytical}
{
\subsection{Planetesimal accretion rate in the statistical model}
In sec.~\ref{sec:comparison}, we use the statistical model for the planetesimal accretion rate presented by \citet{Fortier+2013}.
The planetesimal accretion rate is given by:  \citep[e.g.][]{Chambers2006}
\begin{align}\label{eq:Analitycal_Collision_Rate_detail}
    \frac{d  M_{\rm cap}}{d t} = \frac{2 \pi {R_{\rm H}}^{2}}{P_{\rm orb}} \Sigma_{\rm sol} P_{\rm col},
\end{align}
where $R_{\rm H}$ is the hill radius of the protoplanet, $P_{\rm orb}$ is the orbital period of the protoplanet, and $P_{\rm col}$ is the non-dimensional collision probability.
Planetesimals are in difference velocity regimes depending on their random velocities.
In high-, medium-, and low-velocity regimes, the collision probability is given by: 
\begin{align}
    P_{\rm high} &= \frac{{\tilde{r}}^2}{2 \pi}             \left( I_{\rm F} (\beta) +\frac{6 I_{\rm G} (\beta)}{\tilde{r} {\tilde{e}}^2 } \right), \\
    P_{\rm med}  &= \frac{{\tilde{r}}^2}{4 \pi \tilde{i}}   \left( 17.3 +\frac{232}{\tilde{r}} \right), \\
    P_{\rm low}  &= 11.3 {\tilde{r}}^{1/2},
%    P_{\rm high} &= \frac{{R_{\rm cap}}^2}{2 \pi {R_{\rm H}}^2}             \left( I_{\rm F} (\beta) +\frac{6 R_{\rm H} I_G (\beta)}{R_{\rm cap} {\tilde{e}}^2 } \right), \\
%    P_{\rm med}  &= \frac{{R_{\rm cap}}^2}{4 \pi {R_{\rm H}}^2 \tilde{i}}   \left( 17.3 +\frac{232 R_{\rm H} }{R_{\rm cap}} \right), \\
%    P_{\rm low}  &= 11.3 \left( \frac{R_{\rm cap}}{R_{\rm H}} \right)^{1/2},
\end{align}
where $\tilde{r} \equiv R_{\rm cap}/R_{\rm H}$, and $I_{\rm F}$ and $I_{\rm G}$ are given by: 
\begin{align}
%    \tilde{r}   &= \frac{R_{\rm cap}}{R_{\rm H}}, \\
    I_{\rm F} (\beta) &\simeq \frac{1+0.95925 \beta +0.77251 \beta^2}{\beta (0.13142 +0.12295 \beta)}, \\
    I_{\rm G} (\beta) &\simeq \frac{1 +0.3996 \beta }{\beta (0.0369 +0.048333 \beta +0.006874 \beta^2)},
\end{align}
where $\beta=\tilde{i}/\tilde{e}$.
The mean collision rate can be approximated by  \citep{Inaba+2001}: 
\begin{align}
    P_{\rm col} = {\rm min} (P_{\rm med}, \left( {P_{\rm high}}^{-2} +{P_{\rm low}}^{-2} \right)^{-1/2} ).
\end{align}
Note that we assume that the radius of planetesimals is negligible compared to the capture radius of proto-Jupiter.
$P_{\rm col}$ is expressed as a function of $\tilde{e}$, $\tilde{i}$ and $\tilde{r}$.
The evolution of $\tilde{e}$ and $\tilde{i}$ is determined by the viscous stirring from the protoplanet and the gas drag from the gaseous disk.
The viscous stirring  of planetesimals also increases $\tilde{e}$ and $\tilde{i}$, however, it is neglected in our N-body simulations to save calculation costs.
For consistency, we also neglect the viscous stirring from the planetesimals in the semi-analytical approach.
The rates of the changes in the eccentricity and inclination of planetesimals are given by: 
\begin{align}
    \frac{{\rm d} e^2}{{\rm d} t} &=  \left. \frac{{\rm d} e^2}{{\rm d} t} \right|_{\rm drag} +\left. \frac{{\rm d} e^2}{{\rm d} t} \right|_{\rm VS,M}, \\
    \frac{{\rm d} i^2}{{\rm d} t} &=  \left. \frac{{\rm d} i^2}{{\rm d} t} \right|_{\rm drag} +\left. \frac{{\rm d} i^2}{{\rm d} t} \right|_{\rm VS,M}.
\end{align}
The gas damping rates are given by:  \citep{Adachi+1976,Inaba+2001}
\begin{align}
    \frac{{\rm d} e^2}{{\rm d} t} &= - \frac{2 e^2}{\tau_{\rm aero,0}} \left( \frac{9}{4} \eta^2 +\frac{9}{4\pi} \zeta^2 e^2 +\frac{1}{\pi} i^2 \right)^{1/2}, \\
    \frac{{\rm d} i^2}{{\rm d} t} &= - \frac{  i^2}{\tau_{\rm aero,0}} \left(             \eta^2 +\frac{1}{ \pi} \zeta^2 e^2 +\frac{4}{\pi} i^2 \right)^{1/2},
\end{align}
where $\zeta\sim1.211$ and $\tau_{\rm aero,0}$ is given by: 
\begin{align}
    \tau_{\rm aero,0} &=  \frac{2 m_{\rm pl}}{ C_{\rm d} \pi R_{\rm pl}^2 \rho_{\rm gas} v_{\rm K}}. \label{eq:aerodynamic_gas_drag_timescale0}
\end{align}
The excitation rates of mean square orbital eccentricities and inclinations are given by \citep{Ohtsuki+2002}: 
\begin{align}
    \left. \frac{{\rm d} e^2}{{\rm d} t} \right|_{\rm VS,M} &= \left( \frac{M_{\rm p}}{3 b M_{*} P_{\rm orb}} \right) P_{\rm VS}, \\
    \left. \frac{{\rm d} i^2}{{\rm d} t} \right|_{\rm VS,M} &= \left( \frac{M_{\rm p}}{3 b M_{*} P_{\rm orb}} \right) Q_{\rm VS},
\end{align}
where $b$ is the full width of the feeding zone and is set to  $10$, and $P_{\rm VS}$ and $Q_{\rm VS}$ are given by: 
\begin{align}
    P_{\rm VS} &= \frac{73 {\tilde{e}}^2}{10 \Lambda^2} \ln \left(1 +10\frac{\Lambda^2}{{\tilde{e}}^2} \right)
                    + \frac{72 I_{\rm PVS} (\beta)}{\pi \tilde{e}\tilde{i}} \ln \left(1 +\Lambda^2 \right), \\
    Q_{\rm VS} &= \frac{4 {\tilde{i}}^2 +0.2 \tilde{i}{\tilde{e}}^3 }{10 \Lambda^2 \tilde{e}} \ln \left(1 +10 \Lambda^2 {\tilde{e}} \right)
                    + \frac{72 I_{\rm QVS} (\beta)}{\pi \tilde{e}\tilde{i}} \ln \left(1 +\Lambda^2 \right),
\end{align}
where $\Lambda=\tilde{i} ({\tilde{e}}^2+{\tilde{i}}^2)/12$.
For $0 < \beta \leq 1$, $I_{\rm PVS}$ and $I_{\rm QVS}$ can be approximated by \citep{Chambers2006}: 
\begin{align}
    I_{\rm PVS} (\beta) \simeq \frac{\beta -0.36251}{0.061547 +0.16112 \beta +0.054473 \beta^2}, \\
    I_{\rm QVS} (\beta) \simeq \frac{0.71946 -\beta}{0.21239  +0.49764 \beta +0.14369  \beta^2}.
\end{align}

By solving the above equations and following the evolution of $\tilde{e}$, $\tilde{i}$, we can estimate $P_{\rm col}$.
%$\tilde{r}$ is given by eq.~(\ref{eq:Rplanet_cap}) as in our N-body simulations.

%\subsection{Planetesimal accretion rate in \citet{Shiraishi+2008}}
\subsection{Planetesimal accretion rate presented by Shiraishi \& Ida 2008}
We also use the analytical accretion rate obtained by \citet{Shiraishi+2008}.
When the protoplanet grows rapidly, the expansion speed of the feeding zone regulates the surface density of planetesimals inside the feeding zone.
\citet{Shiraishi+2008} assumed that the accretion rate of planetesimals is regulated by the surface density of planetesimals inside the feeding zone.
Performing the N-body simulations, they derived an analytical expression for the planetesimal accretion rate given by: 
%\begin{align}
%    \frac{d  M_{\rm cap}}{d t} =
%        \begin{cases}
%            \displaystyle{
%                10^{-6} \left( \frac{\rho_{\rm p}}{1 \g/\cm^2} \right)^{1/2} \left( \frac{R_{\rm cap}}{R_\oplus} \right)^2 \left( \frac{\Sigma_{\rm sol}}{2.7 \g/\cm^2} \right) {\eta_{\rm SI}}^{0.8} M_\oplus \yr^{-1}
%                {\rm for}~ \eta_{\rm SI} > 1,
%            } \\
%            \displaystyle{
%                10^{-6} \left( \frac{\rho_{\rm p}}{1 \g/\cm^2} \right)^{1/2} \left( \frac{R_{\rm cap}}{R_\oplus} \right)^2 \left( \frac{\Sigma_{\rm sol}}{2.7 \g/\cm^2} \right) {\zeta_{\rm SI}}^{1.4} M_\oplus \yr^{-1}
%                {\rm for}~ \eta_{\rm SI} < 1,
%            } \\
%        \end{cases} \label{eq:Analytical_Collision_Rate_SI}
%\end{align}
\begin{align}\label{eq:Analytical_Collision_Rate_SI}
    &{\rm for}~ \eta_{\rm SI} > 1; \nonumber \\
    &    \frac{d  M_{\rm cap}}{d t} = 10^{-6} \left( \frac{\rho_{\rm p}}{1 \g/\cm^2} \right)^{1/2} \left( \frac{R_{\rm cap}}{R_\oplus} \right)^2 \left( \frac{\Sigma_{\rm sol}}{2.7 \g/\cm^2} \right) {\eta_{\rm SI}}^{0.8} M_\oplus \yr^{-1}, \\
    &{\rm for}~ \eta_{\rm SI} < 1; \nonumber \\
    &    \frac{d  M_{\rm cap}}{d t} = 10^{-6} \left( \frac{\rho_{\rm p}}{1 \g/\cm^2} \right)^{1/2} \left( \frac{R_{\rm cap}}{R_\oplus} \right)^2 \left( \frac{\Sigma_{\rm sol}}{2.7 \g/\cm^2} \right) {\zeta_{\rm SI}}^{1.4} M_\oplus \yr^{-1},
\end{align}
with
\begin{align}
    \eta_{\rm SI} &\equiv \frac{v_{\rm H}}{v_{\rm scat}} \simeq 4.1 \left( \frac{a_{\rm p}}{5 \AU} \right)^{3/2} \left( \frac{M_{\rm p}}{M_{\oplus}} \right)^{-1/3} \left( \frac{\tau_{\rm acc}}{10^4 \yr} \right)^{-1}, \\
    \zeta_{\rm SI} &\equiv \frac{v_{\rm H}}{v_{\rm aero}} \simeq 0.8 \left( \frac{\tau_{\rm aero}}{10^4 \yr} \right)^{1/2} \left( \frac{M_{\rm p}}{M_{\oplus}} \right)^{-1/6} \left( \frac{a_{\rm p}}{5 \AU} \right)^{3/4},
\end{align}
where $v_{\rm H}$, $v_{\rm scat}$ and $v_{\rm aero}$ are the velocity of the hill sphere expansion, the velocity of the gravitational scattering, and the velocity of the aerodynamic gas damping (see \citet{Shiraishi+2008} for their definition).
%We use the same model as \citet{Fortier+2013} for $\Sigma_{\rm sol}$.

\subsection{Surface density of planetesimals}
In order to infer the planetesimal accretion rate, the surface density of planetesimals must be known. 
In order to calculate $\Sigma_{\rm sol}$, we follow the method developed by \citet{Alibert+2005}, in which the planetesimals are uniformly distributed inside the feeding zone.
$\Sigma_{\rm sol}$ is then given by: 
\begin{align}\label{eq:SurfaceDensityInFeedingZone}
    \Sigma_{\rm sol} = \frac{M_{\rm FZ}}{S_{\rm FZ}},
\end{align}
where $M_{\rm FZ}$ is the total mass of planetesimals inside the feeding zone, and $S_{\rm FZ}$ is the area of the feeding zone.
We consider that the shape of the feeding zone is a ring with a width of $2\sqrt{3} R_{\rm H}$ in the both side of the protoplanet.
$M_{\rm FZ}$ is given by: 
\begin{align}
    M_{\rm FZ} = M_{\rm FZ,in} -M_{\rm cap} -M_{\rm scat},
\end{align}
where $M_{\rm FZ,in}$ is the mass of planetesimals which enter the feeding zone and $M_{\rm scat}$ is the mass of the planetesimals which exits from the feeding zone by the scattering of the protoplanet.
Planetesimals enter the feeding zone crossing both edge of the feeding zone.
The inflow flux of planetesimals $M_{\rm FZ,in}$ is given by: 
\begin{align}
    \frac{{\rm d} M_{\rm FZ,in}}{{\rm d} t} =    2 \pi a_{\rm FZ,out} \dot{a}_{\rm FZ,out} \Sigma_{\rm sol} \left( a_{\rm FZ,out} \right) 
                                                -2 \pi a_{\rm FZ,in} \dot{a}_{\rm FZ,in} \Sigma_{\rm sol} \left( a_{\rm FZ,in} \right).
\end{align}
Planetesimals that are captured and scattered by the protoplanet are removed from the feeding zone.
The scattering rate of planetesimals is give by:  \citep{Ida+2004a}
\begin{align}
    \frac{{\rm d} M_{\rm scat}}{{\rm d} t} &= \left( \frac{v_{\rm esc,p}}{v_{\rm esc,*}} \right)^4 \frac{{\rm d} M_{\rm cap}}{{\rm d} t}, \\
        &= \left( \frac{M_\mathrm{p}}{M_*} \frac{a_\mathrm{p}}{R_\mathrm{cap}} \right)^2 \frac{{\rm d} M_{\rm cap}}{{\rm d} t},
\end{align}
where $v_{\rm esc,p}$ and $v_{\rm esc,*}$ are escape velocity from the planet and the central star.

\section{Additional plots obtained in our study}\label{sec:results_MassiveCore}
%%%%%%%%%%%%%%%%%%%%%%%%%%%%%%%%%%%%%%%%%%%%%%%%%%%%%%%%%%%%%
\begin{figure*}
  \begin{center}
    \includegraphics[width=160mm]{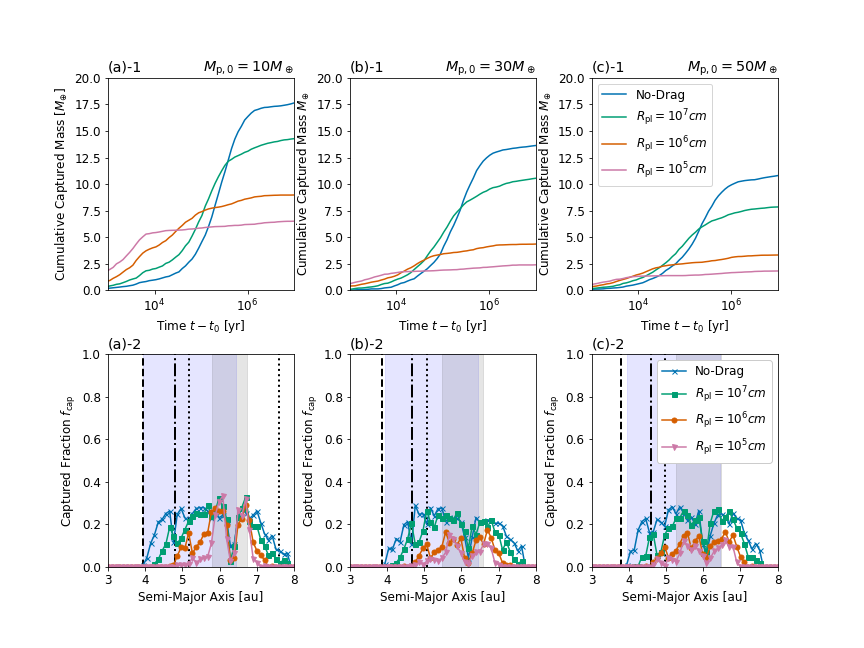}
    \caption{
    The results of numerical simulations where we change the initial mass of proto-Jupiter $M_\mathrm{p,0}$.
    {\bf Upper panel}: We show the cumulative mass of captured planetesimals as a function of the calculation time $t-t_0 [\yr]$.
    {\bf Lower panel}: we show the fraction of the captured planetesimals $f_\mathrm{cap}$ as a function of the initial semi-major axis of planetesimals.
    Left, middle, and right columns show the cases with $M_\mathrm{p,0}=10$, $30$, and $50 M_\oplus$, respectively.
    }
    \label{fig:Results_psMp0}
  \end{center}
\end{figure*}
%%%%%%%%%%%%%%%%%%%%%%%%%%%%%%%%%%%%%%%%%%%%%%%%%%%%%%%%%%%%%
In this section we show the results when we use different initial masses for proto-Jupiter, i.e.,   $M_\mathrm{p,0}$.
Figure~\ref{fig:Results_psMp0} shows the cumulative mass of captured planetesimals $M_\mathrm{cap}$ as a function of the calculation time $t-t_0$, and the fraction of the captured planetesimals $f_\mathrm{cap}$ as a function of the initial semi-major axis of planetesimals.
The total mass of captured planetesimals $M_\mathrm{cap,tot}$ decreases with the increasing $M_\mathrm{p,0}$ because the surface density of planetesimals $\Sigma_\mathrm{solid}$ is smaller due to the more massive core.

%%%%%%%%%%%%%%%%%%%%%%%%%%%%%%%%%%%%%%%%%%%%%%%%%%%%%%%%%%%%%
\begin{figure}
  \begin{center}
    \includegraphics[width=80mm]{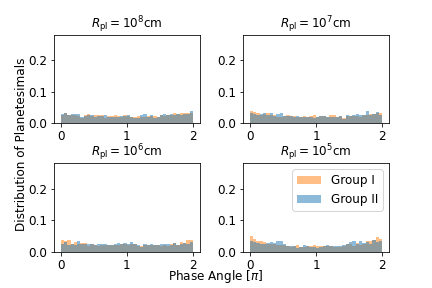}
    \caption{
    Same as panel (b) in fig.~\ref{fig:j0_fcap_woMig}, but show the cases in Model-B and of $\langle {e_0}^2 \rangle^{1/2}=2\langle {\sin }^2 i_0 \rangle^{1/2}=e_\mathrm{eq,m-M}$.
    }
    \label{fig:phase_angle_psRpl}
  \end{center}
\end{figure}
%%%%%%%%%%%%%%%%%%%%%%%%%%%%%%%%%%%%%%%%%%%%%%%%%%%%%%%%%%%%%

Figure~\ref{fig:phase_angle_psRpl} shows the distribution of phase angles in the cases of $\langle {e_0}^2 \rangle^{1/2}=2\langle {\sin }^2 i_0 \rangle^{1/2}=e_\mathrm{eq,m-M}$.
We find that almost all the  planetesimals are outside the mean motion resonances suggesting that the effect of MMRs would be negligible in these cases.

}

\bibliography{refs}{}
\bibliographystyle{mnras}

\bsp	% typesetting comment
\label{lastpage}

\end{document}